\documentclass[10pt]{elsart}
\input{target_tracker.sty}

\begin{document}
%\linenumbers \pagewiselinenumbers

\begin{frontmatter}

\title{The OPERA experiment Target Tracker}

\author[ires]{T.~Adam},
\author[neuchatel]{E.~Baussan},
\author[bern]{K.~Borer},
\author[lal]{J-E.~Campagne},
\author[ires]{N.~Chon-Sen},
\author[lal]{C.~de~La~Taille},
\author[ires]{N.~Dick},
\author[ires]{M.~Dracos\thanksref{contact}},
\author[ires]{G.~Gaudiot},
\author[ires]{T.~Goeltzenlichter},
\author[dubna]{Y.~Gornushkin},
\author[ires]{J-N.~Grapton},
\author[ires]{J-L.~Guyonnet},
\author[bern]{M.~Hess},
\author[ires]{R.~Igersheim},
\author[neuchatel]{J.~Janicsko~Csathy},
\author[ires]{C.~Jollet},
\author[neuchatel]{F.~Juget},
\author[ires]{H.~Kocher},
\author[dubna]{A.~Krasnoperov},
\author[dubna]{Z.~Krumstein},
\author[lal]{G.~Martin-Chassard},
\author[bern]{U.~Moser},
\author[dubna]{A.~Nozdrin},
\author[dubna]{A.~Olchevski},
\author[dubna]{S.~Porokhovoi},
\author[lal]{L.~Raux},
\author[dubna]{A.~Sadovski},
\author[ires]{J.~Schuler},
\author[bern]{H-U.~Sch\"{u}tz},
\author[ires]{C.~Schwab},
\author[dubna]{A.~Smolnikov},
\author[ulb]{G.~Van~Beek},
\author[ulb]{P.~Vilain},
\author[bern]{T.~W\"{a}lchli},
\author[ulb]{G.~Wilquet},
\author[ires]{J.~Wurtz}

\address[bern]{University of Bern, CH-3012 Bern, Switzerland}
\address[ulb]{IIHE- Universit\'{e} Libre de Bruxelles, 1050 Brussels, Belgium}
\address[dubna]{ JINR-Joint Institute for Nuclear Research, 141980 Dubna, Russia}
\address[neuchatel]{Universit\'{e} de Neuch\^{a}tel, CH 2000 Neuch\^{a}tel, Switzerland}
\address[lal]{LAL, Universit\'{e} Paris-Sud 11, CNRS/IN2P3, Orsay, France}
\address[ires]{IPHC, Universit\'{e} Louis Pasteur, CNRS/IN2P3, Strasbourg, France}

\thanks[contact]{Corresponding author, marcos.dracos@ires.in2p3.fr}

\begin{abstract}
The main task of the Target Tracker detector of the long baseline neutrino
oscillation OPERA experiment is to locate in which of the target
elementary constituents, the lead/emulsion bricks, the neutrino
interactions have occurred and also to give calorimetric information
about each event. The technology used consists in walls of two
planes of plastic scintillator strips, one per transverse direction.
Wavelength shifting fibres collect the light signal emitted by the
scintillator strips and guide it to both ends where it is read by
multi-anode photomultiplier tubes. All the elements
used in the construction of this detector and its main characteristics are described.
\end{abstract}

\begin{keyword}
OPERA, Target Tracker, plastic scintillator, photomultiplier, PMT, multianode,
WLS fiber
\end{keyword}

\end{frontmatter}

\section{Introduction} \label{introduction}

OPERA~\cite{opera} is a long baseline neutrino oscillation
experiment designed to detect the appearance of $\nu_\tau$ in a pure
$\nu_\mu$ beam in the parameter region indicated by the anomaly in
the atmospheric neutrino flux.
The detector is installed at the Laboratori Nazionali
del Gran Sasso (LNGS) in a cavern excavated under the Gran Sasso in
the Italian Abruzzes. The cavern (Hall C) is in the line of sight of
the CNGS beam of neutrinos originating from CERN, Geneva, at a
distance of 730~km. The beam energy has been optimized to maximize
the number of $\nu_\tau$ CC interactions. The commissioning of the
beam and of the electronic components of the detector has started
in August 2006. The data taking is due to start in
spring 2007 and to last for five years.

At the nominal value of $\Delta m^2_{23}=2.5\ 10^{-3}$ eV$^2$ and
full $\nu_\mu-\nu_\tau$ mixing ($\sin^22\theta_{23}=1$), as measured
by the Super--Kamiokande atmospheric neutrino
experiment~\cite{superk}, OPERA will observe about 20 $\nu_\tau$ CC
interactions after 5 years of data taking, with an estimated
background of only 1 event.

The OPERA detector consists of two identical super--modules.
Each super--module has a 0.9-kton instrumented target followed by a
$10\times 8$~m$^2$ dipolar magnetic muon spectrometer. One target is the
repetition of 31 $6.7\times 6.7$~m$^2$ modules each including a
proper target wall followed by a tracker wall. A target wall is an
assembly of 52 horizontal trays, each of which is loaded with 64 bricks of
8.3~kg each. A brick is made of 56 lead sheets, 1~mm thick,
providing the necessary mass, interleaved with 57 nuclear emulsion
films that provide the necessary sub--millimeter spatial resolution
required to detect and separate unambiguously the production and
decay vertices of the $\tau^-$ lepton produced in charged current
$\nu_\tau$ interactions with the lead nuclei.

The bricks in which neutrino interactions have occurred, typically
30 per day at the nominal beam intensity, are identified by the
event reconstruction in the trackers and the spectrometers. They
will be extracted on a regular basis, disassembled and the emulsion
films will be scanned and analyzed by a battery of high speed, high
resolution automatic microscopes in order to locate the interaction
vertex and search for candidates of $\tau^-$ lepton decay.

The main role of the Target Tracker is therefore to locate the
lead/emulsion brick where a neutrino interaction has occurred. It will
also provide a neutrino interaction trigger for the readout of the
whole OPERA detector and be used as a calorimeter for the event
analysis.

The required high brick finding efficiency puts strong requirements
on the Target Tracker spatial resolution and track detection
efficiency. The replacement of faulty elements of the Target Tracker
is extremely difficult and this detector must therefore present a
long term stability and reliability (at least for 5 years which is
the expected OPERA data taking period). In case of problems, the
brick finding efficiency not only of the bricks just in front of the
concerned zone, but also of several walls upstream will severely be
affected. The sensitive surface to be covered being of the order of
$2\times 3000$~m$^2$, a cost effective technology had to be used.

\section{Overview of the Target Tracker} \label{overview}

\begin{figure}[hbt]
\begin{minipage}{.45\linewidth}
\begin{center}
\mbox{\epsfig{file=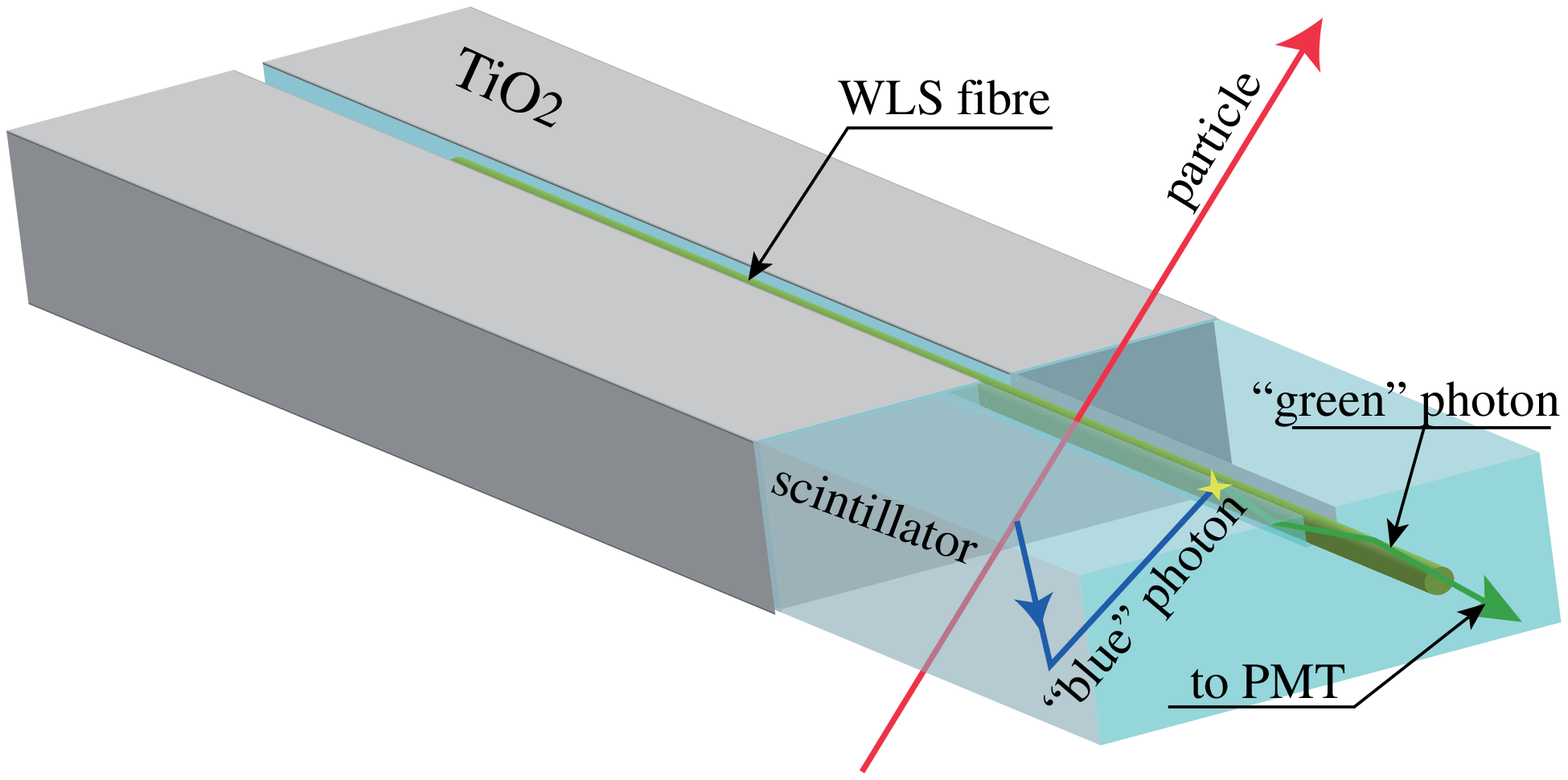,width=7cm}}
\caption{\small Particle detection principle in a scintillating
strip.}\label{principle}
\end{center}
\end{minipage} \hspace{1.cm}
\begin{minipage}{.45\linewidth}
\begin{center}
\mbox{\epsfig{file=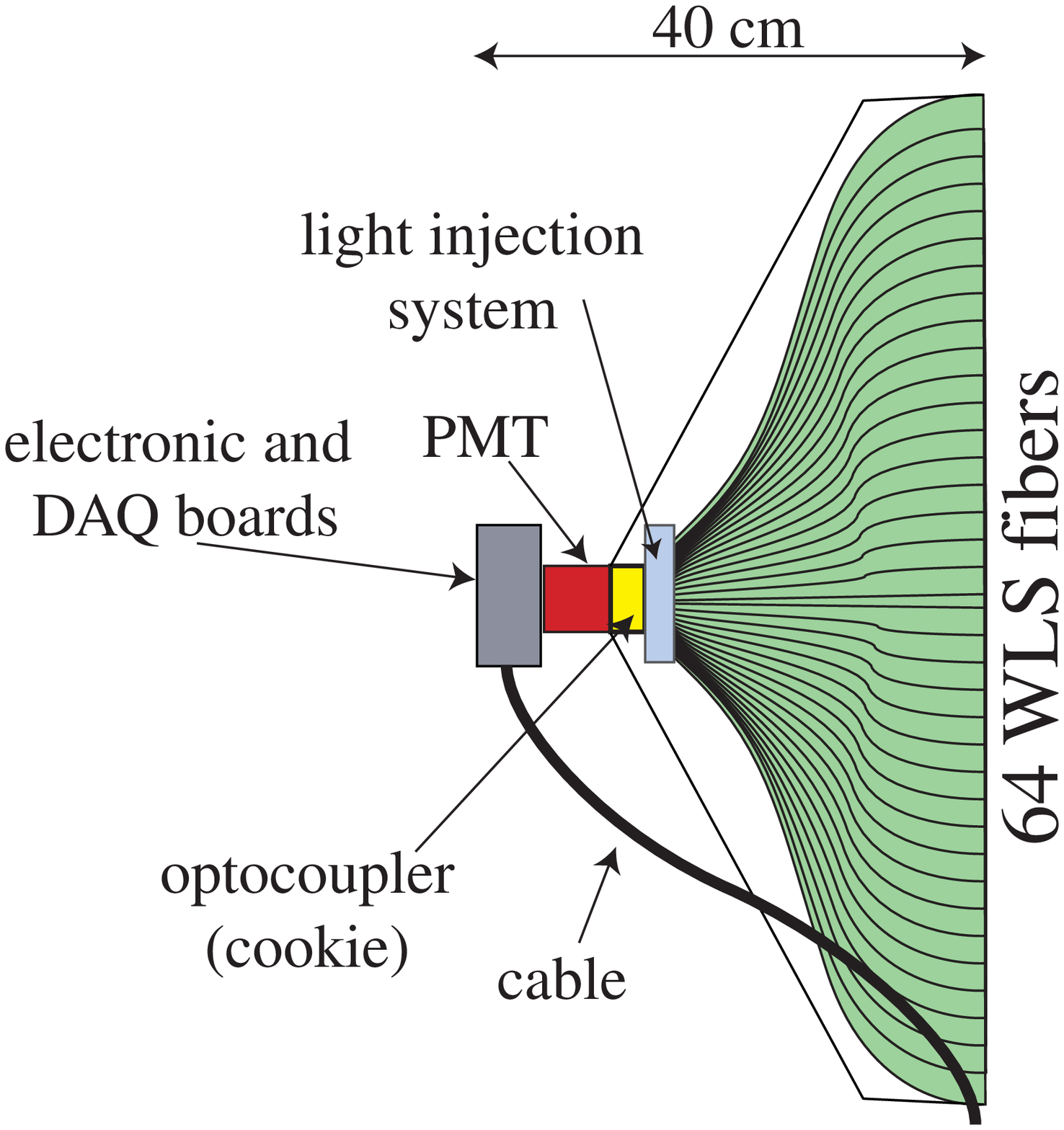,width=7cm}}
\caption{\small Schematic view of an end--cap of a scintillator
strip module.}\label{endcap_schematic}
\end{center}
\end{minipage}
\end{figure}

The technology selected to instrument the targets of the OPERA
detector consists in scintillator strips, 6.86~m long, 10.6~mm
thick, 26.3~mm wide, read on both sides using Wave Length Shifting
(WLS) fibres and multi-anode photomultipliers (PMT). The particle
detection principle is depicted by Fig.~\ref{principle}. The
scintillator strips used have been produced by extrusion, with a
$TiO_2$ co-extruded reflective and diffusing coating for better
light collection. A long groove running
on the whole length and at the center of the scintillating strips,
houses the WLS fibre which is glued inside the groove using a high
transparency glue. This technology is very reliable due to the
robustness of its components. Delicate elements, like electronics
and PMT's are located outside the sensitive area where they are accessible.

A basic unit (module) of the target tracker consists of 64 strips
readout by WLS fibres coupled to two 64--channel photodetectors.
The 64 strips of the module are glued on the surrounding aluminium sheets
that serve as covers (the strips are not glued together).
In this way, the mechanical strength of the module is provided by the strips
themselves, the covers and the end--caps.
The fibres are directly routed at both ends to the
photodetectors through the end--caps (Fig.~\ref{endcap_schematic}).

\begin{figure}[hbt]
\begin{minipage}{.45\linewidth}
\begin{center}
\mbox{\epsfig{file=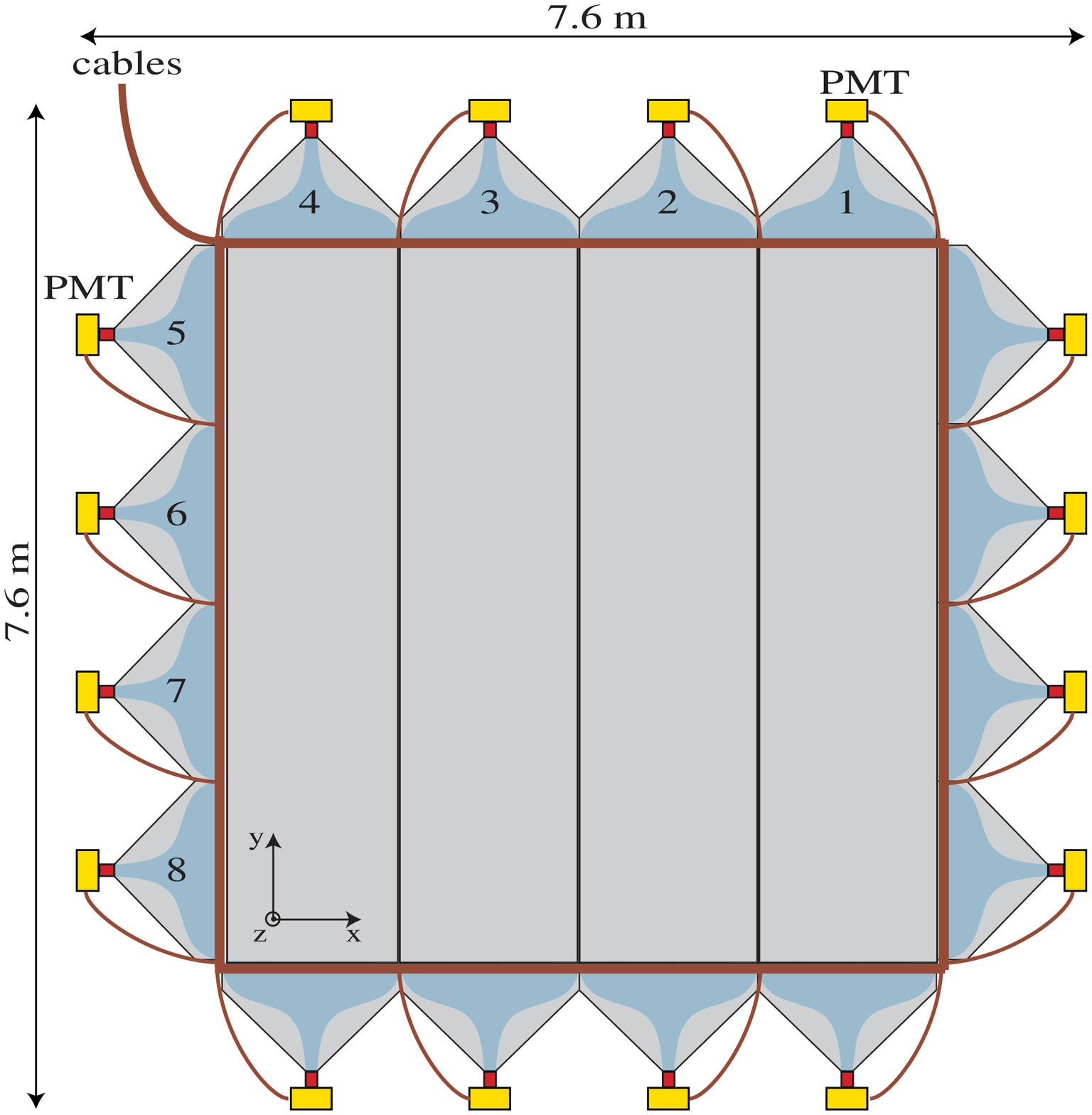,width=7cm}}
\caption{\small Schematic view of a plastic scintillator strip
wall.} \label{wall_schematic}
\end{center}
\end{minipage} \hspace{1.cm}
\begin{minipage}{.45\linewidth}
\begin{center}
\mbox{\epsfig{file=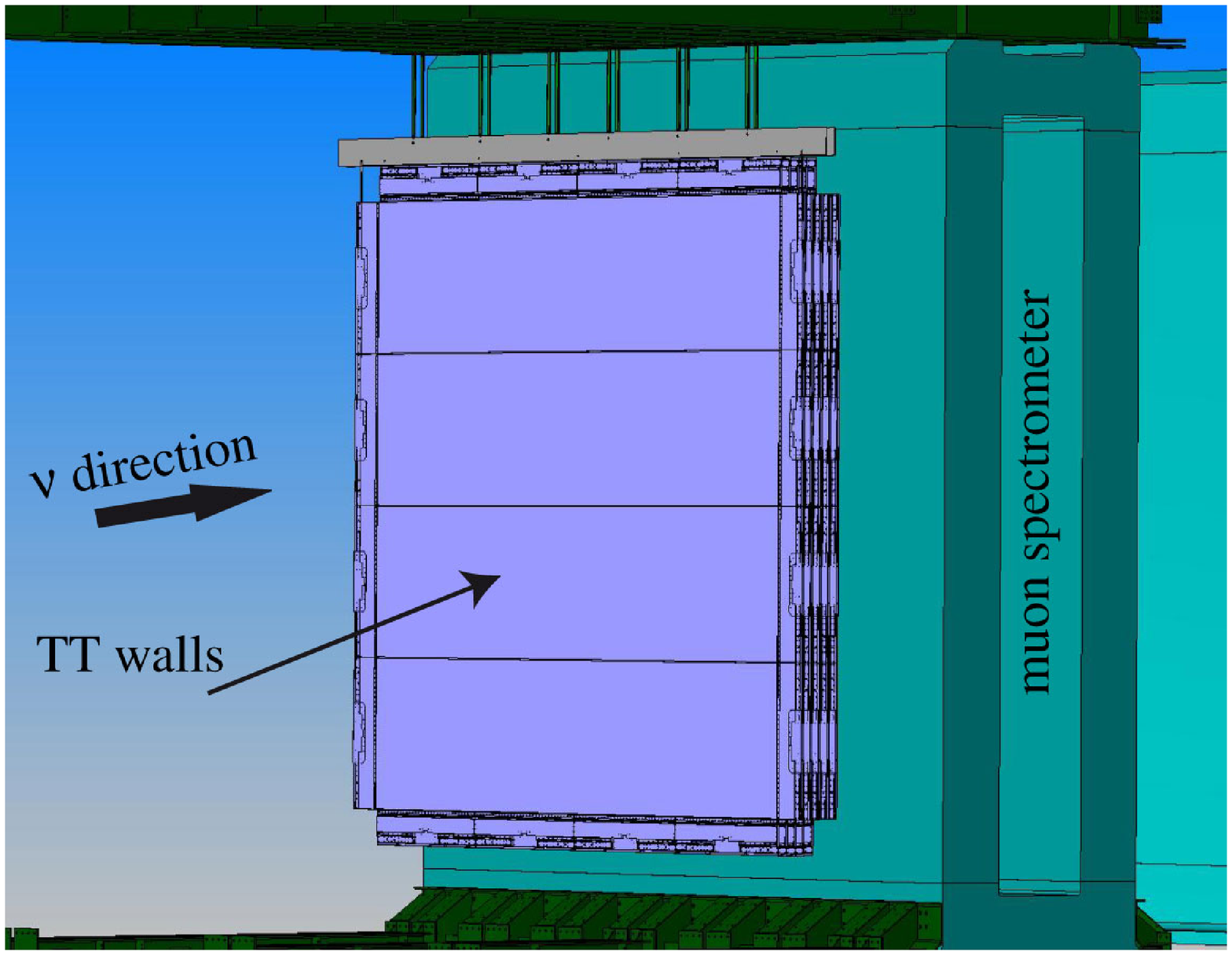,width=7cm}}
\caption{\small Target Tracker walls hanging in between two brick
walls inside the OPERA detector.} \label{hanging}
\end{center}
\end{minipage}
\end{figure}

Four such modules are assembled together to construct a tracker plane
covering the $6.7\times 6.7$~m$^2$ sensitive surface defined by the
target brick walls. One plane of 4 horizontal modules and one plane
of 4 vertical modules form a tracker wall providing 2D
track information (Fig.~\ref{wall_schematic}). The total OPERA
target contains 62 walls, 31 per super--module. Thus, the total number of
scintillating strips is 31744 for 63488 electronic channels.

Target Tracker walls hanging in between two brick walls are shown
schematically by Fig.~\ref{hanging}, the walls are suspended on
the OPERA main I--beams independently of the brick walls. The
vertical modules are suspended on a flat beam (10~mm thick,
30~cm wide and 8.24~m long) which is hanged directly on the
main I--beams of the OPERA detector supporting structure,
while for the horizontal modules only the first one
is suspended at the two ends on the flat beam while the other
three modules are suspended on one another. The end--caps of the
horizontal modules are rigid enough to support the other
modules.

The minimum total wall thickness in the sensitive area covered by
the emulsion bricks is 28.8~mm. The dead space induced by the gap
between modules, the mechanical tolerance between strips of 0.1~mm
and the scintillator strip $TiO_2$ wrapping (0.15~mm thick) is of
the order of 1.5\% of the overall sensitive surface.

\section{Components} \label{components}

A description of the main components entering the Target Tracker
construction is given in this section. More information can be found
in the Target Tracker Technical Design Report~\cite{TTTDR}.

\subsection{Plastic Scintillator Strips} \label{scintillators}

\begin{figure}[hbt]
\begin{minipage}{.45\linewidth}
\begin{center}
\mbox{\epsfig{file=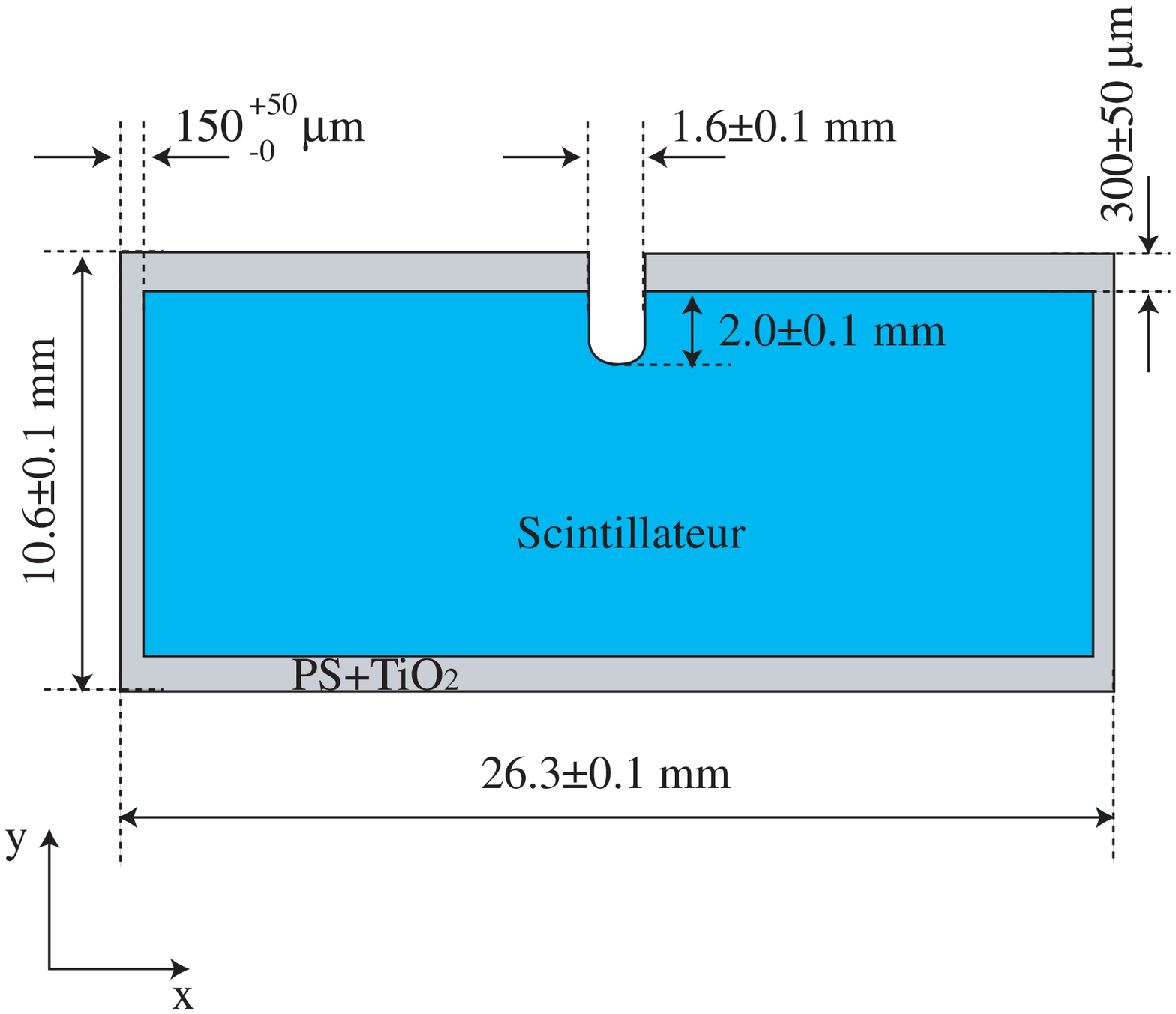,width=7cm}}
\caption{\small Scintillator strip geometry (total strip
length=$6860^{+0}_{-2}$~mm at 20$^\circ$C).} \label{strip_geo}
\end{center}
\end{minipage} \hspace{1.cm}
\begin{minipage}{.45\linewidth}
\begin{center}
\mbox{\epsfig{file=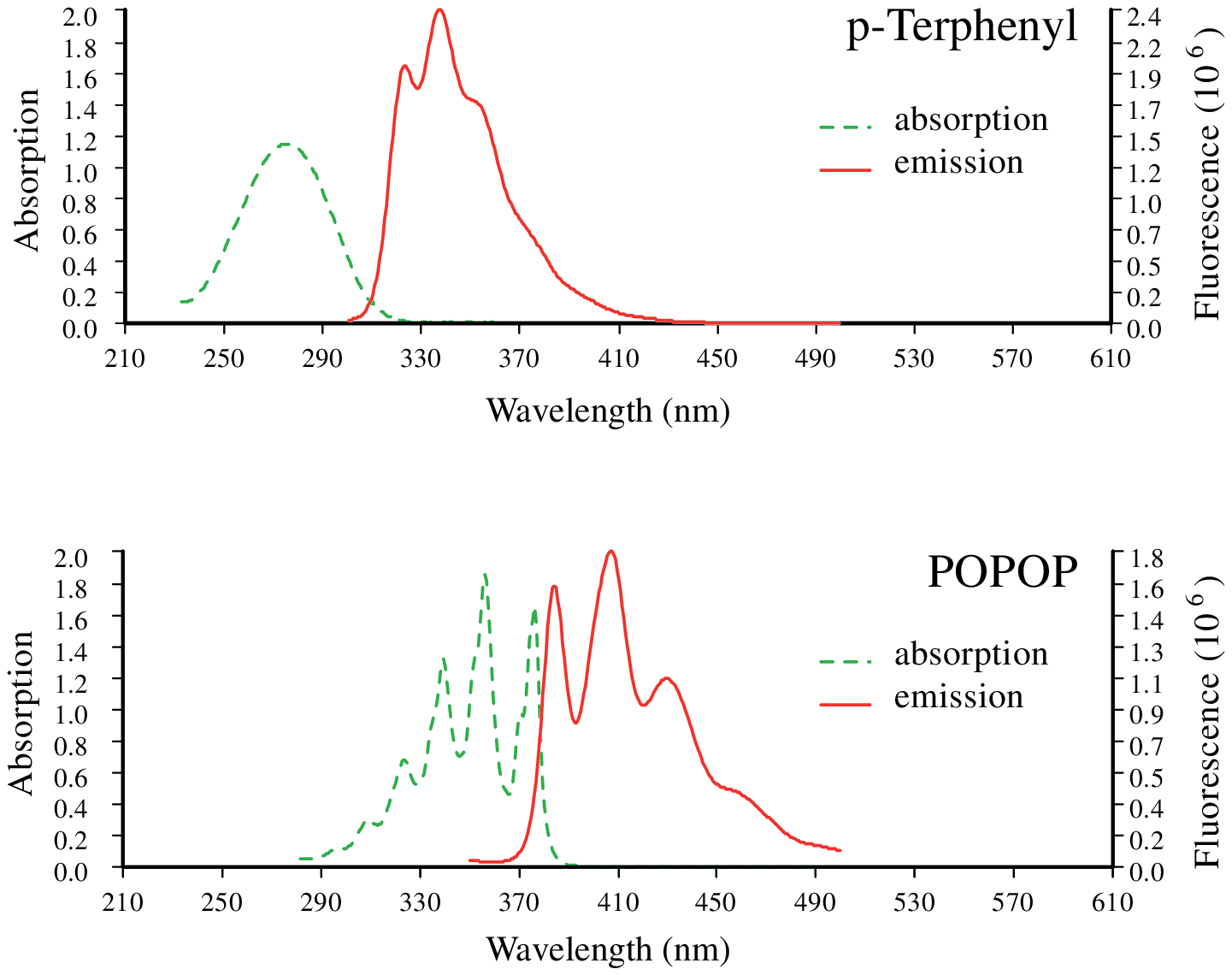,width=7cm}}
\caption{\small Absorption and emission spectra of primary
(p-Terphenyl) and secondary (POPOP) fluors.}
\label{absemi}\end{center}
\end{minipage}
\end{figure}

The transverse geometry of the plastic scintillator strips is shown
in Fig.~\ref{strip_geo}. The scintillator strips have been extruded,
with a co-extruded $TiO_2$ reflective coating for improved light
collection, by the AMCRYS-H company\footnote{AMCRYS-H, 60, Lenin
ave, Kharkov, 310001, Ukraine.}. The WLS fibre is glued with high
transparency glue in a machined groove, 2.0~mm deep and 1.6~mm wide,
which runs along the strip length. The plastic scintillator is
composed of polystyrene with 2\% of p-Terphenyl, the primary fluor,
and 0.02\% of POPOP, the secondary fluor. Fig.~\ref{absemi}~\cite{photoch} shows
the absorption and emission spectra of the fluors.

\begin{figure}[hbt]
\begin{minipage}{.45\linewidth}
\begin{center}
\mbox{\epsfig{file=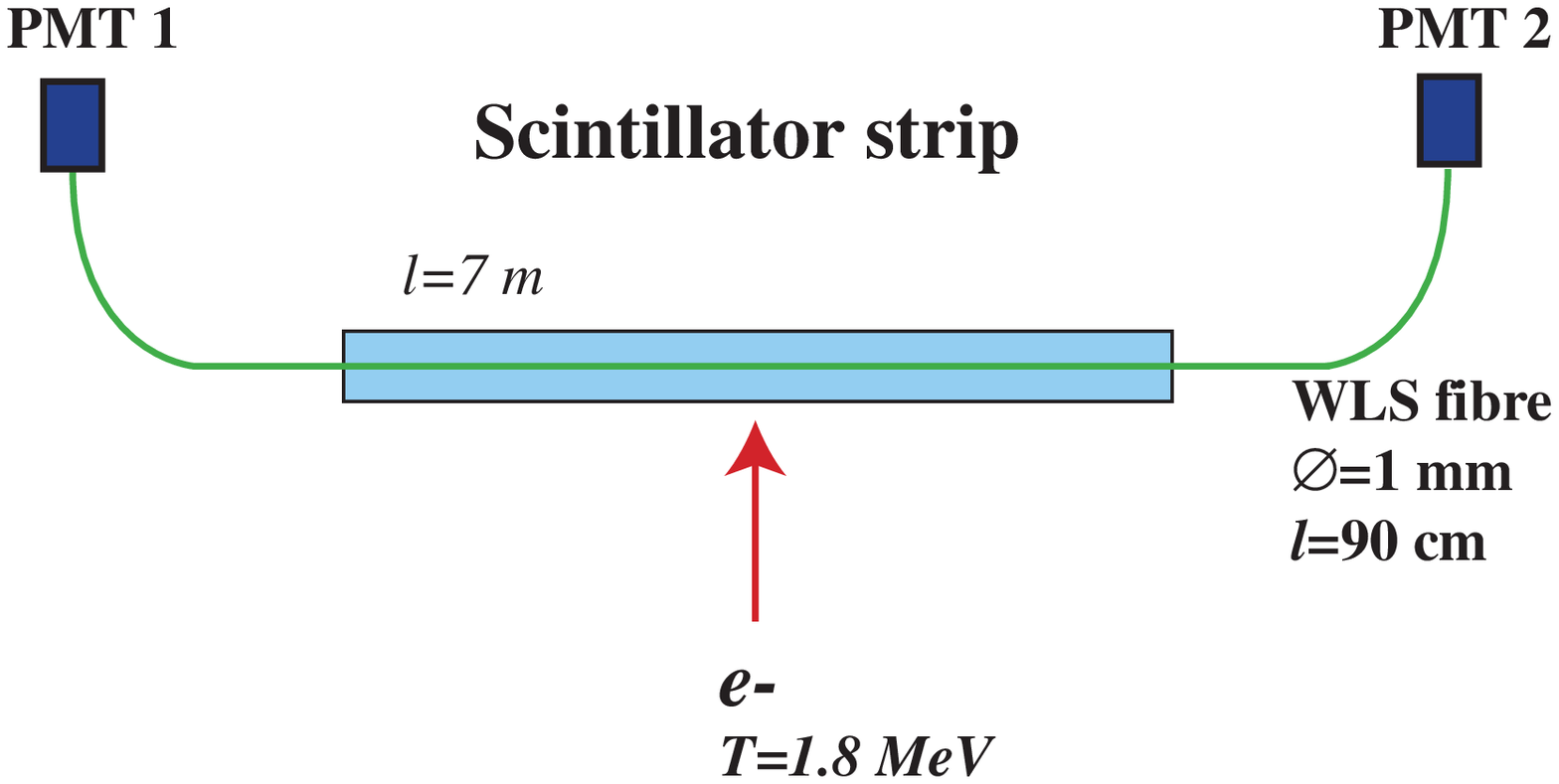,width=7cm}}
\caption{\small Setup used for scintillator comparison using an
electron spectrometer.}\label{scint_test}
\end{center}
\end{minipage} \hspace{1.cm}
\begin{minipage}{.45\linewidth}
\begin{center}
\mbox{\epsfig{file=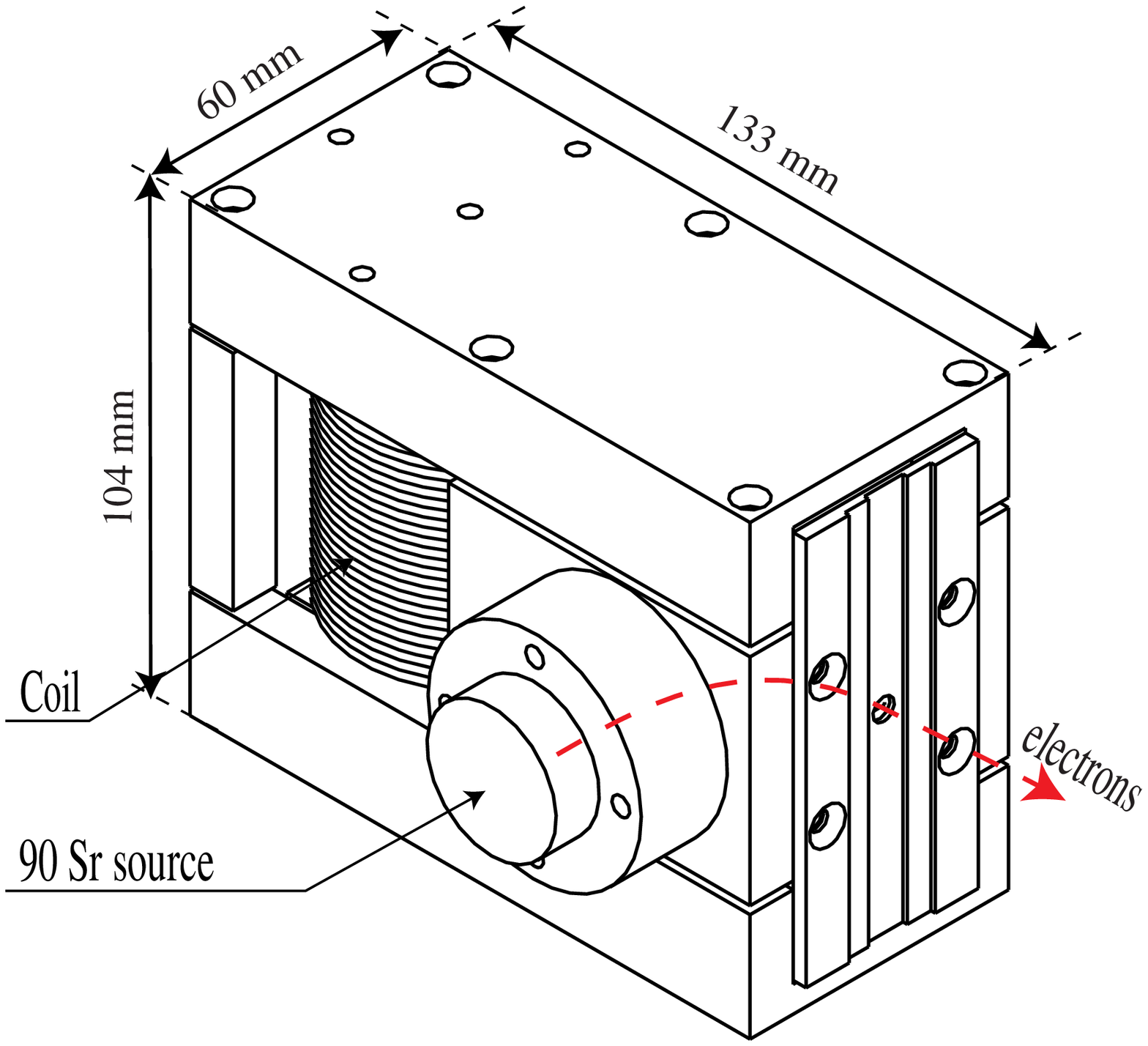,width=7cm}}
\caption{\small Electron spectrometer.} \label{spectro}
\end{center}
\end{minipage}
\end{figure}

Several plastic scintillator strips produced by different companies
have been tested by irradiation of the strips
(Fig.~\ref{scint_test}) with electrons of 1.8~MeV energy selected
from a 10~mCi $^{90}Sr$ $\beta$ source by a purposely developed compact magnetic
spectrometer (Fig.~\ref{spectro}). The electron trigger at the exit
of the spectrometer is provided by a 0.1~mm thick scintillator read
at both ends by two PMT's in coincidence.
The signal was read-out by
two Hamamatsu bialkali PMT's H3164-10 through the WLS fibre.
A correction factor
of  1.19 was applied to the measurements to express the signal in
terms of the number of photoelectrons (p.e) resulting from the
2.15~MeV energy lost by a crossing minimum ionizing particle.
Subsequent tests with a pion beam at the CERN-PS have confirmed this
correction factor within 4\%.

\begin{figure}[hbt]
\begin{minipage}{.45\linewidth}
\begin{center}
\mbox{\epsfig{file=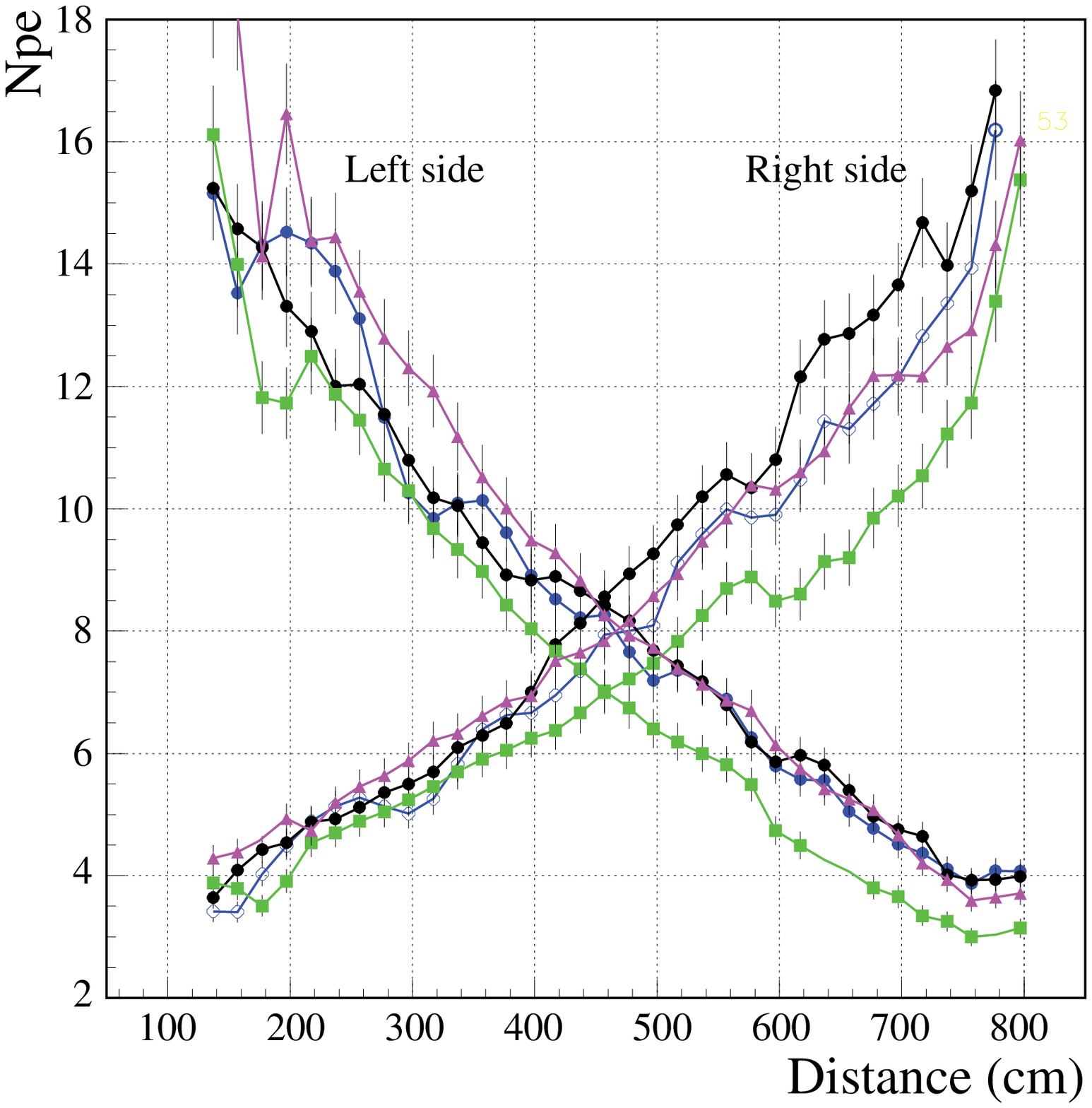,width=7cm}}
\caption{\small Number of p.e versus the distance from
the PMT's for AMCRYS-H strip samples. For this measurement, Kuraray
Y11 (175) fibres have been used.}\label{amcrys_npe}
\end{center}
\end{minipage} \hspace{1.cm}
\begin{minipage}{.45\linewidth}
\begin{center}
\mbox{\epsfig{file=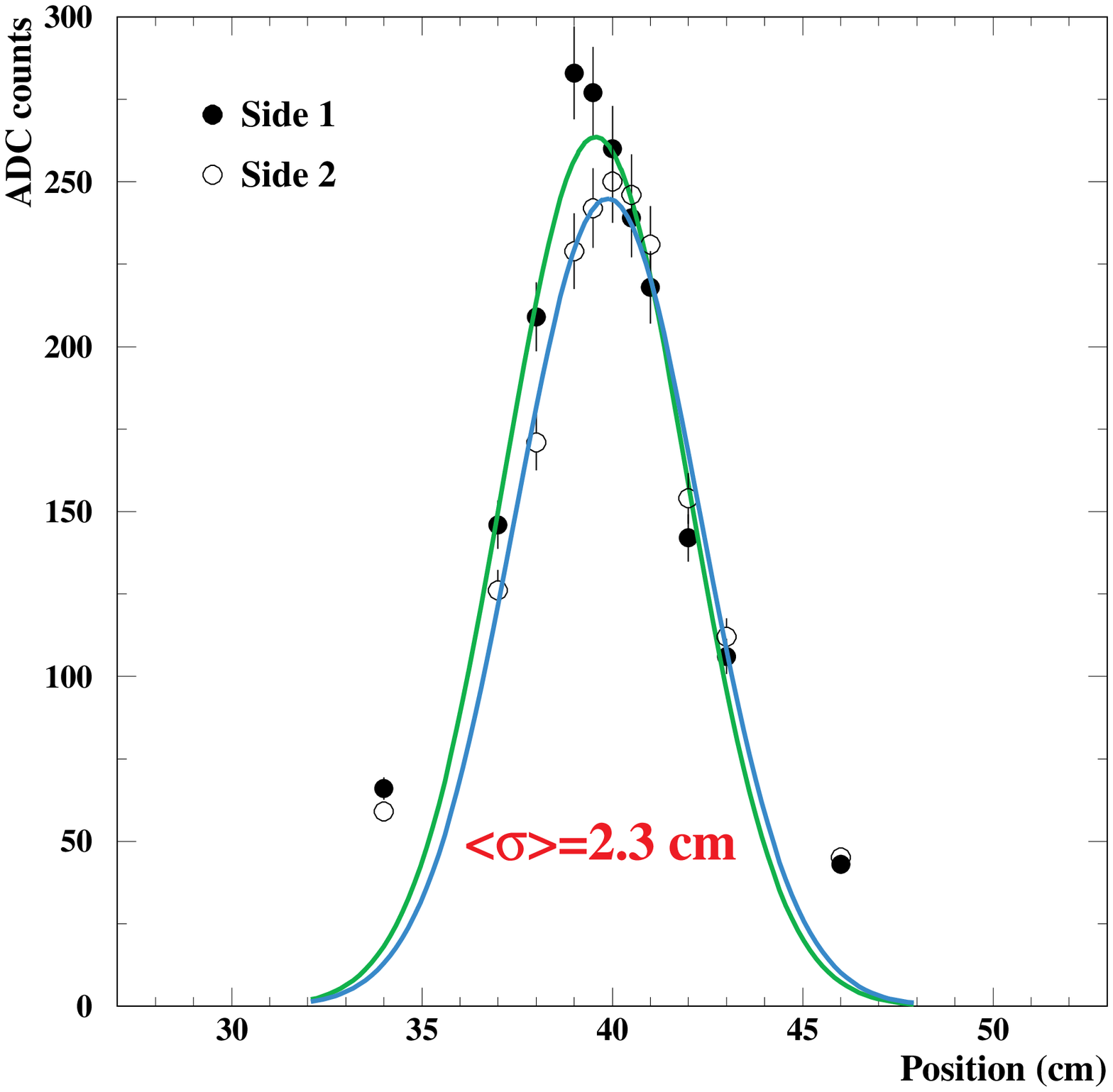,width=6cm}}
\caption{\small Light collection versus the position of the WLS
fibre piece.} \label{light_collection}
\end{center}
\end{minipage}
\end{figure}

Fig.~\ref{amcrys_npe} presents the number of p.e
versus the distance from the two PMT's for several samples
provided by AMCRYS-H company. In the worst case where the particle
crosses the strip at its middle (4.5~m distance from each PMT),
the number of observed p.e is well above 4 (lower limit requested by the experiment), inducing a particle
detection efficiency higher than 99\%.

Primary photons produced in the strip by an ionizing particle make
several reflections on the strip surface before entering the WLS
fibre. To evaluate the distribution of the distance between the
point of emission and the entry point in the fibre, the long WLS
fibre was replaced by a short segment of 3~cm glued at both ends to
clear fibres. By varying the position of the spectrometer with
respect to the center of the WLS fibre segment, the measured light
yield shown by Fig.~\ref{light_collection} was observed. A gaussian
distribution has been fitted to the data with a standard deviation of 2.3~cm.
Taking into account the WLS fibre length, the standard deviation of the light
expansion distribution has been calculated to be about 2.2~cm.

%\clearpage
\subsection{Wavelength Shifting Fibres}

\begin{figure}[hbt]
\begin{center}
\mbox{\epsfig{file=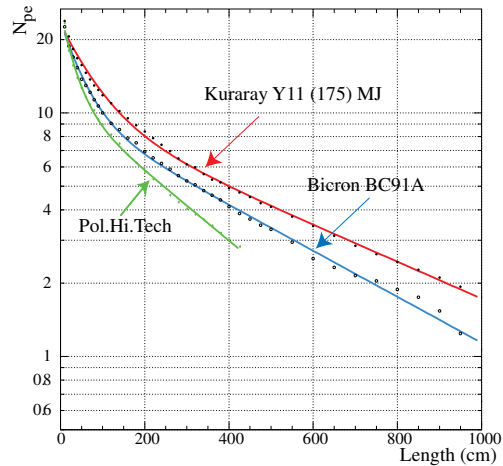,width=7cm}}
\caption{\small Fibre attenuation using 1~mm diameter Kuraray,
Bicron and Pol.Hi.Tech fibres.}\label{fibre_atten}
\end{center}
\end{figure}

The attenuation length has been measured for several 1~mm diameter
double cladding WLS fibres commercially available from
Bicron\footnote{Bicron Corp., 12345 Kinsman Road, Newbury, Ohio
44065.}, Kuraray\footnote{Kuraray Co., Methacrylic Resin Division,
8F, Maruzen Building, 3-10, 2-Chrome, Hihonbashi, Chuo-ku, Tokyo,
103-0027, Japan.} and PoL.Hi.Tech\footnote{Pol.Hi.Tech. s.r.l.,
67061 CARSOLI (AQ), S.P. Turanense Km. 44.400, Italy.}. The fibres
were inserted into a 1~mm diameter hole machined in a NE110
scintillator excited by an $H_2$ UV lamp. Fig.~\ref{fibre_atten}
shows the collected light intensity versus the distance between the
PMT (Hamamatsu bialkali H3164-10) and the illumination point for the
Y11(175)MJ non S fibre from Kuraray, which was the final choice for
this detector. The equivalent number of p.e is just
indicative. The fitted curve is the sum of two exponential
distributions: $e^{2.59-x/\lambda_s}+e^{2.29-x/\lambda_l}$ with
$\lambda_s=79$~cm (short absorption length) and $\lambda_l=573$~cm (long absorption length).

%\clearpage
\subsection{End-Caps} \label{end-caps}

\begin{figure}[hbt]
\begin{center}
\epsfig{file=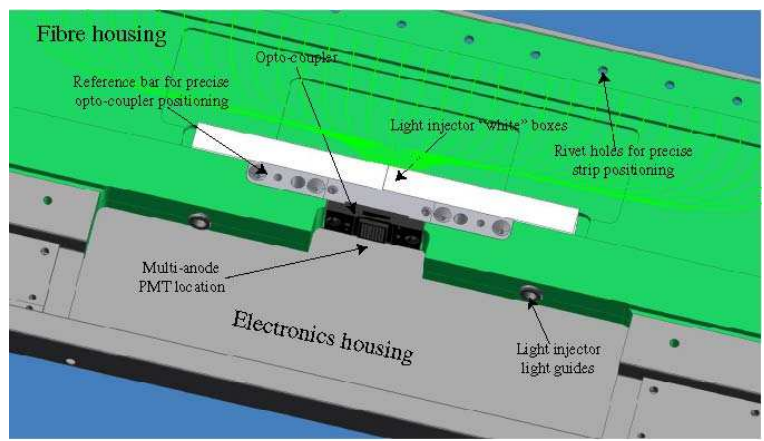,width=10cm}
\caption{\small 3D view of the central part of an end--cap.}
\label{endcap_3D}
\end{center}
\end{figure}

The fibres are routed to the PMT photocathodes in the two module
end-caps (Fig.~\ref{endcap_3D}). These also constitute the
mechanical structure by or to which the modules are suspended in the
OPERA detector. The length of the end-cap, 1.7~m, is defined by the module width.
Its width is fixed by the minimum fibre bending radius, the
photodetector and its opto--coupling window and the electronics.
Altogether it equals the maximum value of 40~cm, a limit imposed by
the design of the manipulator used for brick insertion into and
extraction from the target walls. It has a maximum thickness of
3.4~cm and weighs about 10~kg. The core of the body is a block of
320~kg/m$^3$ polyurethane foam glued on a black soft steel frame.
Soft steel has been chosen for the sake of shielding the PMT from
the spectrometer fringe magnetic field.

The housing for the free ends of the optical fibres is machined into the
foam body. It is lined with soft black tissue for the sake
of protecting the fibre cladding from scratches and increasing the
light tightness. Every other strip is attached to one end-cap,
and so on for the other 32. For this, the end-cap is equipped with 32
rivets of 7.3~mm diameter spaced by 52.8~mm, twice the maximum
scintillator strip width plus 0.1~mm tolerance between strips. Each
rivet is due to receive the hole drilled at one end of a strip. The
fibre housing is closed by the optical coupler providing a precise
positioning of each fibre in front of its corresponding PMT channel.
The frame also provides housing for the multi-channel PMT and its HV
power supply, the monitoring light injection system, the
front--end electronics and the data acquisition cards. The low
voltage and readout cables are extracted and routed through a path
on the back of the end-caps. When closed with its main cover, the
light tightness in the fibre housing is ensured.

The end--caps have been constructed by the A\'eriane
company\footnote{A\'eriane S.A., rue des Poiriers, 7, B-5030
Gembloux, Belgium.}.

%\clearpage
\subsection{Photodetectors} \label{photodetector}

The choice of the photodetector is mainly based on the single
p.e detection efficiency, the dynamic range, the cost
and the geometry. Other considerations are gain uniformity among
channels, linearity and cross--talk. The photodetector chosen for
the OPERA Target Tracker is based on the commercially available
$64$-channel Hamamatsu H$7546$ PMT. These PMT's are powered by a
negative polarity high voltage (in the following, the absolute
value of the high voltage will be given). This PMT has also been
chosen for the MINOS~\cite{minos} near detector and has been
extensively evaluated.

\begin{figure}[hbt]
\begin{center}
\epsfig{file=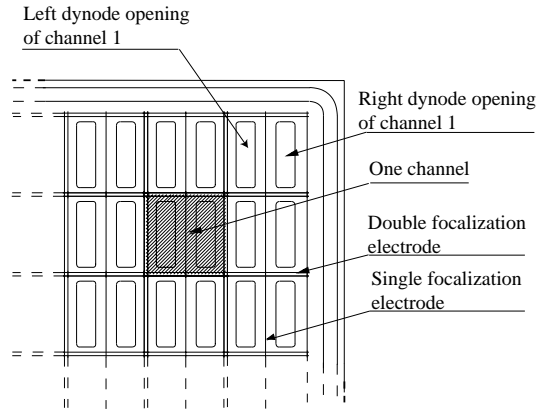,width=7cm} \caption{\small Schematic
view of a part of a multianode PMT showing the channel separation by
focalization electrodes.} \label{electrodes}
\end{center}
\end{figure}

Each channel contains two sets of 12 dynodes and covers a
surface of $2.3\times 2.3$~mm$^2$ (Fig.~\ref{electrodes}). The PMT
provides a common output of the signals on the last dynodes which can
be used as a FAST-OR to trigger the
acquisition system or for timing purposes.

\begin{figure}[hbt]
\begin{center}
\epsfig{file=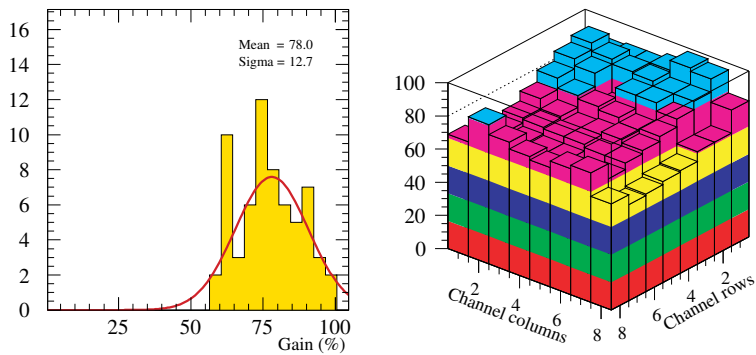,width=10cm}
\caption{\small Left: distribution of the responses of the 64
channels of a Hamamatsu H7546 PMT's at 800~V, Right: lego--plot of
the responses on the photocathode area. The signal is normalized to
100.} \label{pmtuni}
\end{center}
\end{figure}

Fig.~\ref{pmtuni} is an example of the uniformity of the channels
responses, normalized to 100 and obtained by full photocathode
illumination using a W--lamp~\cite{hamamatsu}. More advanced studies
of the multi--anode PMT properties have been performed on a
purposely developed test bench. A computer guided translation system
and the PMT are enclosed in a light--tight box together with a $H_2$
UV lamp, a band--pass filter and focalization optics. The full area
of the photocathode can be scanned with a narrow light spot with a
diameter $<50~\mu m$. Alternatively, the photocathode may be
illuminated by a WLS fibre glued to a bloc of scintillator activated
by the UV lamp.

\begin{figure}[hbt]
\begin{minipage}{.45\linewidth}
\begin{center}
\mbox{\epsfig{file=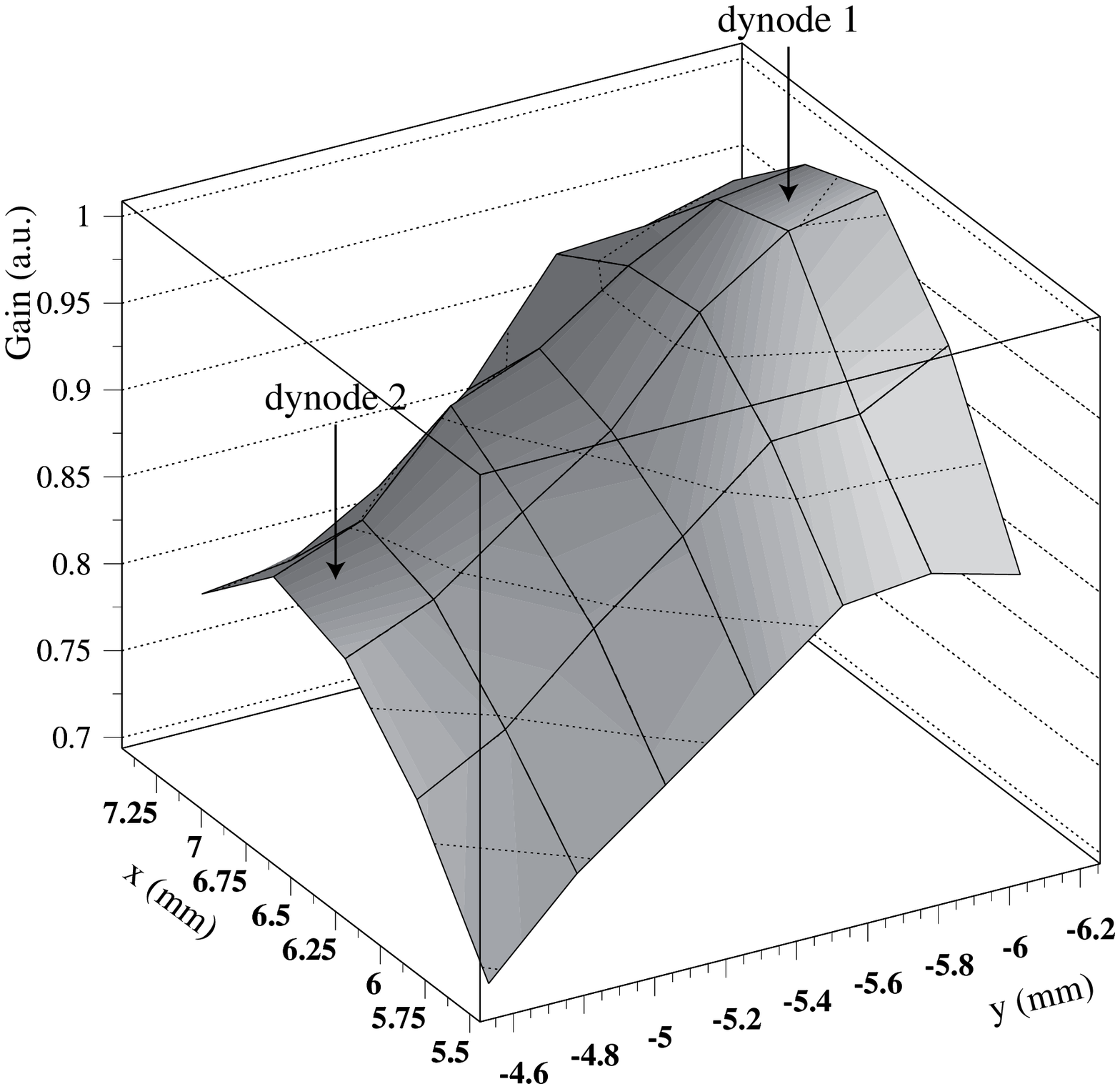,width=7cm}}
\caption{\small Gain variation over a single PMT cell
(channel).}\label{scangain}
\end{center}
\end{minipage} \hspace{1.cm}
\begin{minipage}{.45\linewidth}
\begin{center}
\mbox{\epsfig{file=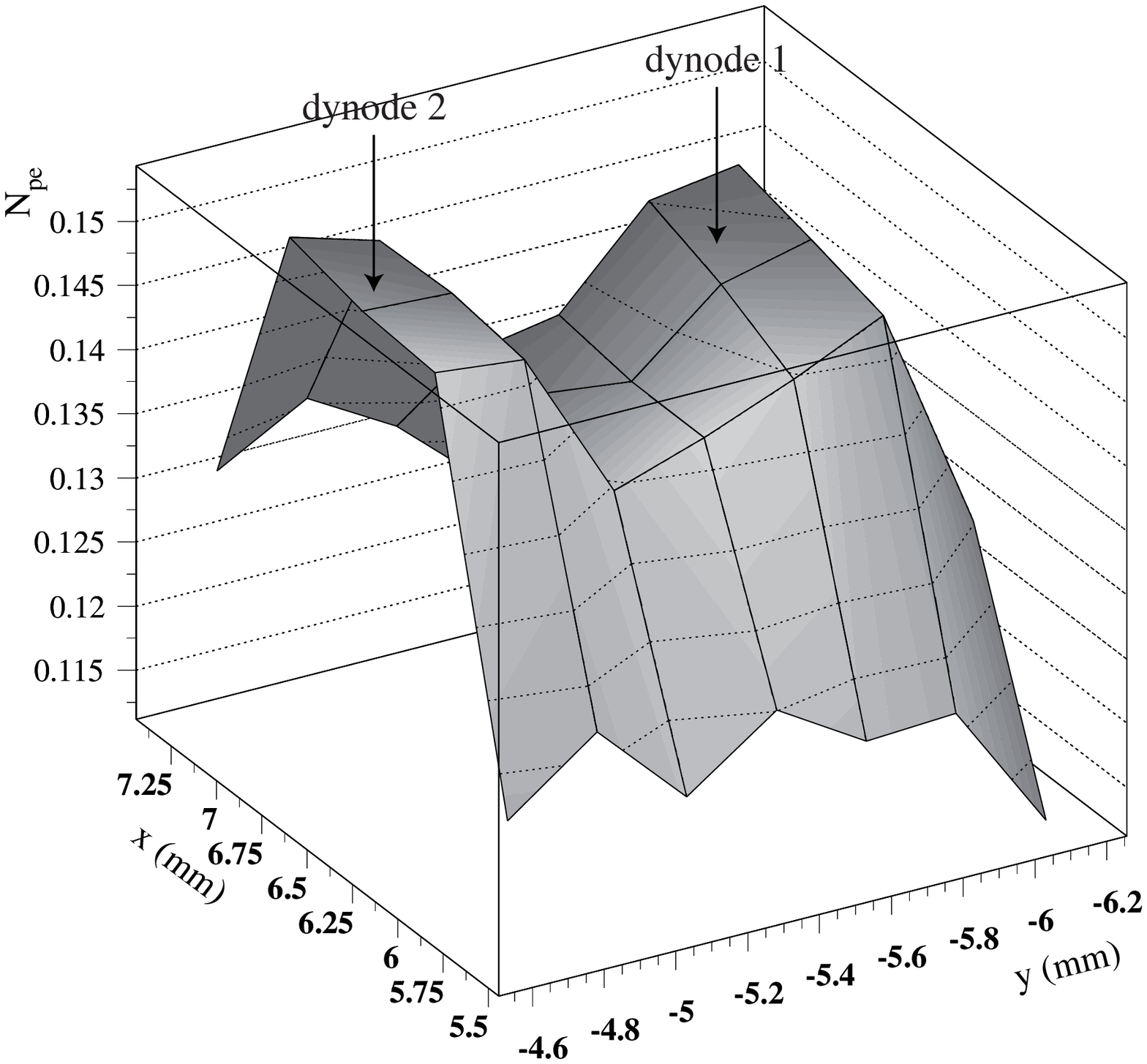,width=7cm}}
\caption{\small Distribution of the number of p.e
collected over a single PMT cell.} \label{scannpe}
\end{center}
\end{minipage}
\end{figure}

\begin{figure}[hbt]
\begin{minipage}{.45\linewidth}
\begin{center}
\mbox{\epsfig{file=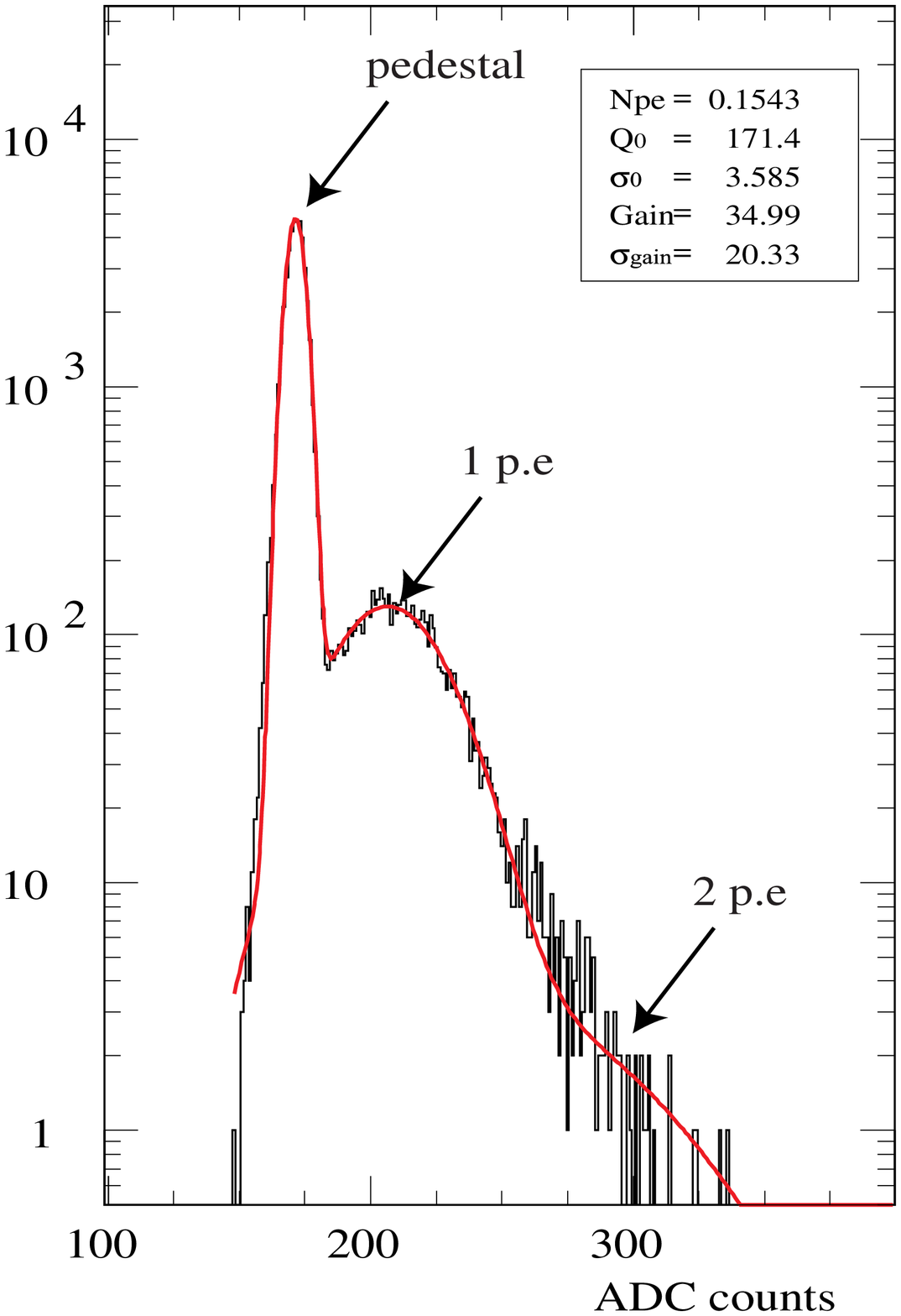,width=5cm}}
\caption{\small Charge distribution recorded by one PMT
channel.}\label{scanone}
\end{center}
\end{minipage} \hspace{1.cm}
\begin{minipage}{.45\linewidth}
\begin{center}
\mbox{\epsfig{file=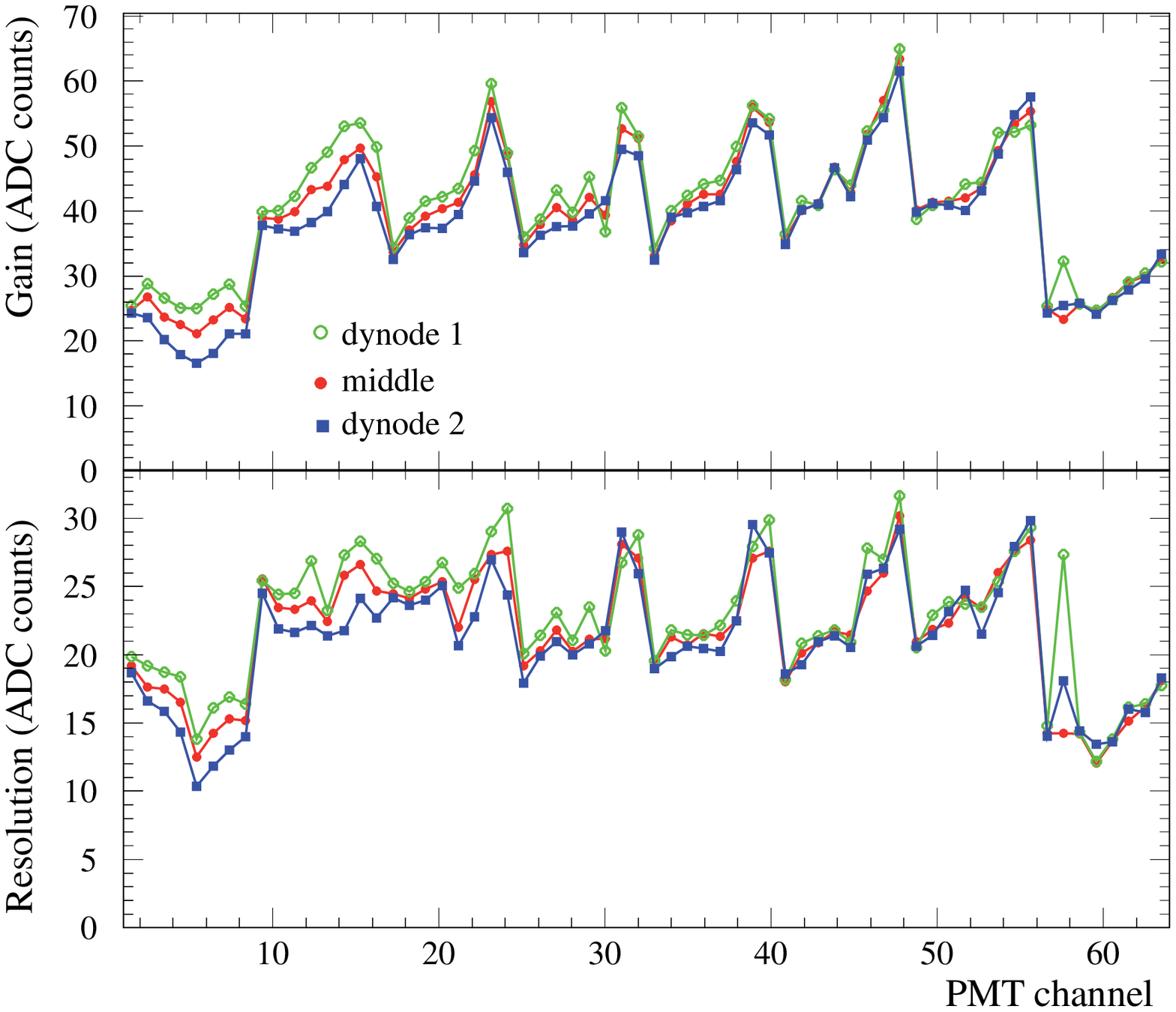,width=7cm}}
\caption{\small Gain and resolution of each PMT channel when the
fibre is at the middle of the channel and in front of each dynode.}
\label{gd}
\end{center}
\end{minipage}
\end{figure}

Fine scans of the gain distribution and of the number of p.e
collected inside a single channel are shown on Fig.~\ref{scangain}
and Fig.~\ref{scannpe}. They were obtained with a 1.0~mm fibre and a
light intensity reduced to the single p.e level. A loss of
p.e collection of the order of 6\% is observed when the
fibre is positioned between the two dynodes as is in the experiment.
The number of p.e was computed by fitting the charge
distribution with the convolution of a gaussian and a Poisson
distribution for the signal and a gaussian distribution for the
pedestal (Fig.~\ref{scanone})~\cite{bellamy}. The maximum dispersion
between the two dynodes of the same channel is of the order of 20\%
while the gain variation from channel to channel may reach a factor
3 (Fig.~\ref{gd}). The dispersion on the gain is about 50\% of the gain mean value of each channel.

The counting rate induced by natural radioactivity or PMT dark
current must be as low as possible to reduce the effect of dead
time. Using the PMT with a threshold corresponding to $1/3$ of
p.e, a noise of less than 10~Hz/channel at $25^o$C has
been measured coming from photocathode thermo--emission. This
possibility of using a low threshold ensures a very high single p.e
detection efficiency.

The PMT's have been customized by Hamamatsu to our requests.
Modifications had to do with mechanics (assembly and alignment of
the tube with the optical coupler), electronics (rearrangement of the
back plane connectors) and light tightness (injection of black, high
voltage resistant resin in the space between the tube and its
housing).

\begin{figure}[hbt]
\begin{center}
\epsfig{file=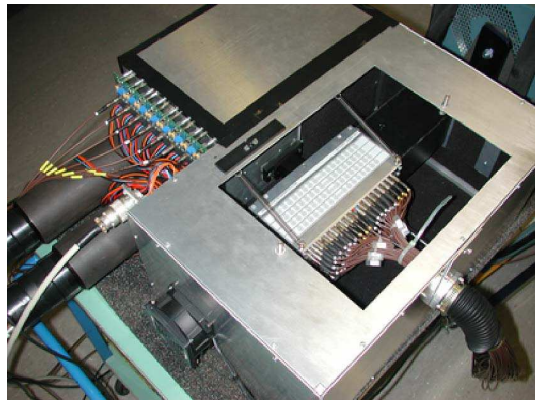,width=7cm} \caption{\small The multianode PMT
testing device.} \label{setuppix}
\end{center}
\end{figure}

All PMT's were connected to a reference optical coupling window and
passed a number of calibration measurements (\cite{thomas}). These
were achieved with a device in which the 64 WLS--fibres of the
coupling window were divided into 8 groups of 8 arranged such that
no neighbouring fibres belong to the same group. Each group of
fibres originated in a separate light diffusion box containing two
UV-LED's SLOAN L5-UV5N\footnote{SLOAN AG, Birmannsgasse 8, CH-4009
Basel/Switzerland.}. The fibres were exposed over a length of 2~mm to
light pulses, the intensity of which could be varied by a factor of
about 250. The position of the two LED's with respect to the fibres
was chosen such as to extend the dynamic range of the system from
0.03 to 120 p.e. The signals of two additional monitoring
fibres arranged across each box were read by two reference Hamamatsu
HI949 PMT's. The monitoring of the LED light output with time was
provided by comparison to the light emitted by a piece of scintillator loaded with
a weak 30~Bq Bi source. The testing device is shown in
Fig.~\ref{setuppix} where the amplifier box developed especially for
this application is seen.

One of the goals of the PMT tests was to determine the nominal value
of the high voltage to be applied to each unit. The high voltage was
defined such that the gain of the strongest channel of each PMT
equals $10^6$. The mean high voltage applied on all tested PMT's is
825~V varying from 750~V to 930~V. The gains of the 64 channels are
electronically equalized by means of the front end chips
(see~\ref{electronics}). Along with the gain, other important
characteristics were measured, such as cross--talk and
dark current. All relevant results were stored in a data base for
use during the calibration of the Target Tracker modules.

To avoid problems along time, it has been decided not to apply
optical grease between the opto--coupler and the PMT photocathode.
Optical grease may also generate bubbles between the PMT
and opto--coupler surface which could create light transmission problems
and also increase the optical cross--talk between channels.
This choice causes a loss of about 15\%
of the number of observed p.e.

For the high voltage power supply of the PMT's, small modules
located for convenience on the DAQ boards in the end--caps, have
been selected. The modules provided by Iseg company\footnote{Iseg
Spezialelektronik GmbH, Bautzner Landstr. 23, D - 01454
Radeberg/Rossendorf.} (BPS BPN 10 165 12) fulfill the following main
requirements:

\begin{itemize}
\item Adjustable voltage (negative polarity) between 0 and 1000~V
by means of an external control voltage not exceeding 5~V.
\item Accuracy of the output voltage of ±1\%, ripple less than 0.01\% peak-to-peak and
temperature coefficient not exceeding ±0.01\% $^{\circ}$C.
\item  Output maximum current between 1~mA and 2~mA (overload and short circuit protected).
\item Modules powered by a low voltage DC supply (14~V$\pm 10$\%).
\item MTBF (Mean Time Between Failure) higher than
300'000 hours at full load and 25 $^{\circ}$C (a burn--in procedure
during 24 hours at maximum charge with temperature cycling was done
by the company for each module).
\end{itemize}

\begin{figure}[hbt]
\begin{minipage}{.45\linewidth}
\begin{center}
\mbox{\epsfig{file=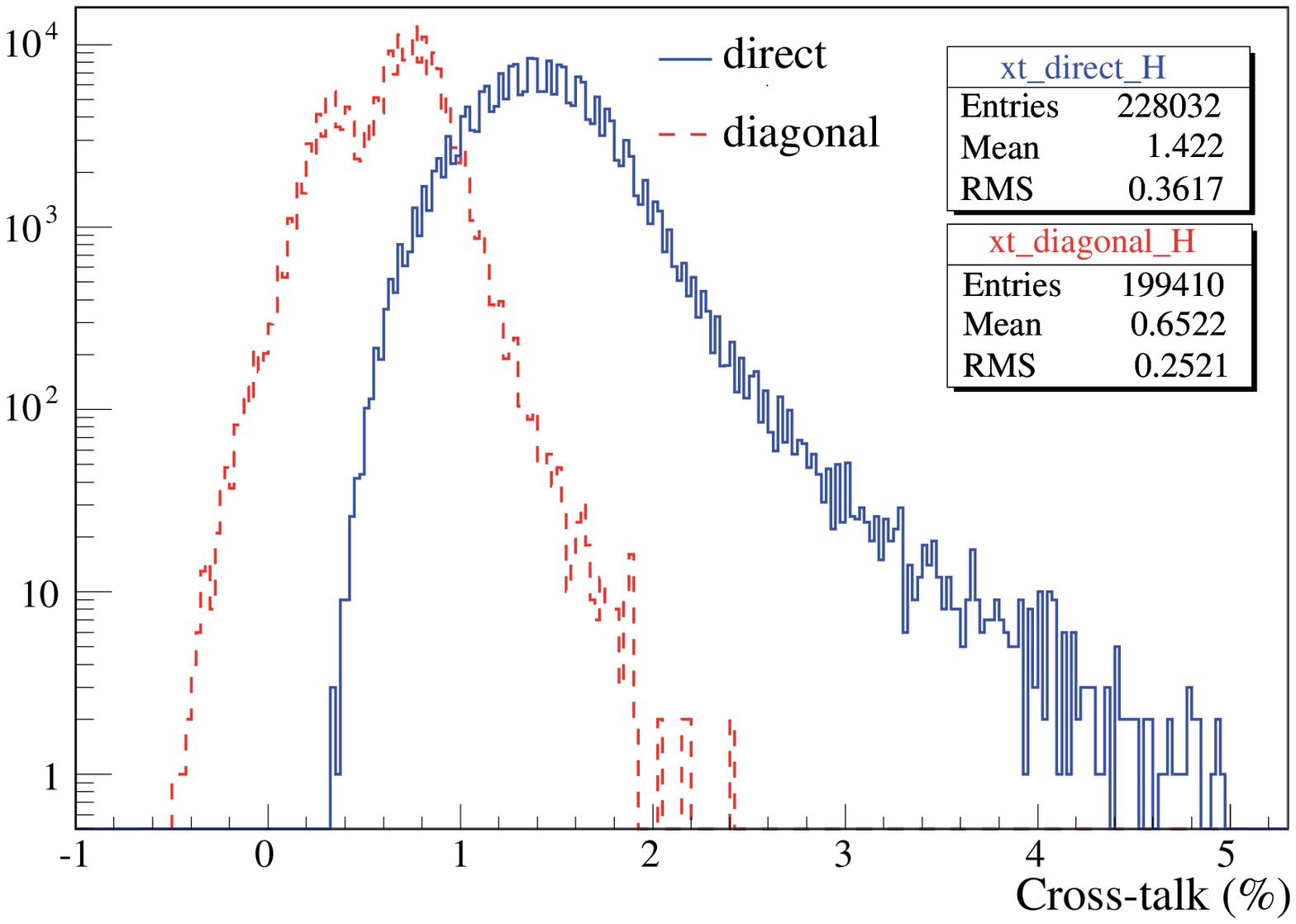,width=7cm}}
\caption{\small Cross--talk distributions for direct and diagonal
neigbours for all tested PMT's and channels.}\label{crosstalk}
\end{center}
\end{minipage} \hspace{1.cm}
\begin{minipage}{.45\linewidth}
\begin{center}
\mbox{\epsfig{file=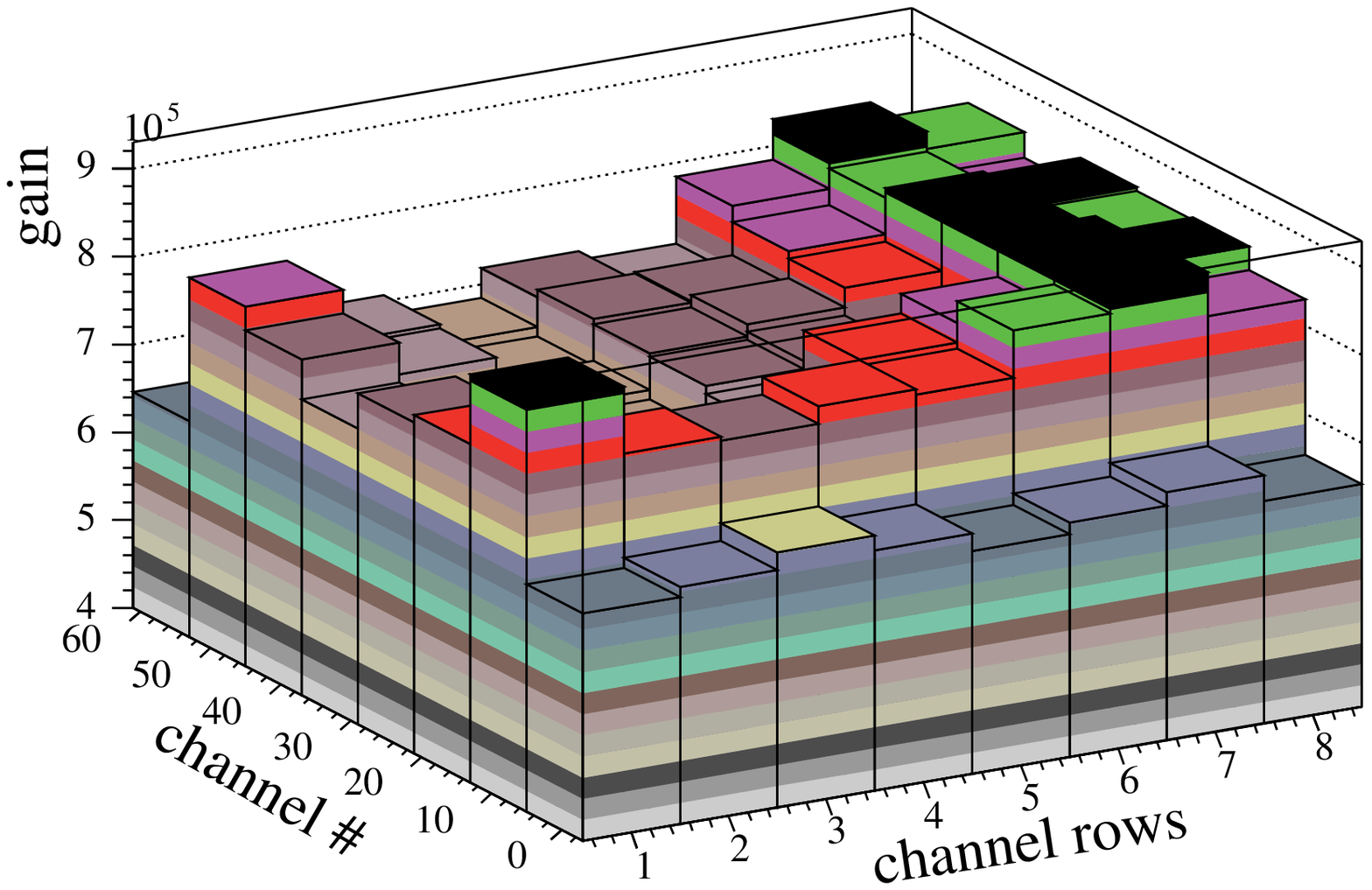,width=7cm}}
\caption{\small Gain versus the channel position in multi--anode
PMT's.} \label{gainall}
\end{center}
\end{minipage}
\end{figure}

After testing 1032 multi-anode PMT's, 5.4\% have been rejected
mainly due to high cross--talk between neighbouring channels. The
cross--talk distributions for direct and diagonal neighbours are
shown in Fig.~\ref{crosstalk}. The mean cross-talk on direct
neighbours is of the order of 1.43\% while for diagonal neighbours
this factor goes down to 0.65\%. Fig.~\ref{gainall} presents the
mean gain of all tested PMT's versus the channel position. One can
see that the channels of the first (1--8) and last (57--64) columns
have a significantly lower gain than the other channels. PMT's where
the difference between the highest and lowest observed gain on
individual channels exceeds a factor of 3 were rejected.

\begin{figure}[hbt]
\begin{center}
\epsfig{file=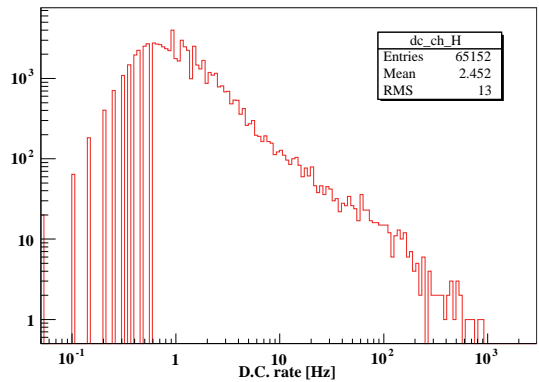,width=7cm} \caption{\small Dark
count rate for all tested PMT channels.} \label{dcrate}
\end{center}
\end{figure}

Another important parameter, especially for OPERA which is a
triggerless experiment, is the PMT dark count rate.
Fig.~\ref{dcrate} presents the dark count rate of all tested PMT
channels. A mean value of 2.45~Hz is observed at 20$^\circ$C. PMT's
having channels with rate higher than 300~Hz were rejected.

%\clearpage
\subsection{Front-End Electronics} \label{electronics}

The readout electronics of the Target Tracker is based on a
32-channel ASIC, referenced in the following as the OPERA ReadOut
Chip (ROC). Two ROC's are used to readout each multi--anode PMT, for
a total of 1984 chips for the full detector. A detailed description
of the ASIC design and performance can be found in \cite{roc}.

The main requirements that have driven the chip design are:

\begin{itemize}
  \item Compensation for the factor 3 anode-to-anode gain variations (Fig.~\ref{gd}).
  It is equipped with an adjustable gain system incorporated in the preamplifier
  stage that delivers a signal of identical range to the fast and slow shapers of every channel.
  \item Delivery of a global low noise auto-trigger and time information with 100\%
  trigger efficiency for particles at minimum of ionization (MIP),
  that is for a signal as low as 1/3 of p.e, corresponding to 50~fC at the anode for a PMT gain of 10$^6$.
  \item Delivery of a charge proportional to the signal delivered by each
  pixel of the PMT in a dynamic range corresponding to 1 to 100 p.e.
\end{itemize}

\begin{figure}[hbt]
\begin{center}
\epsfig{file=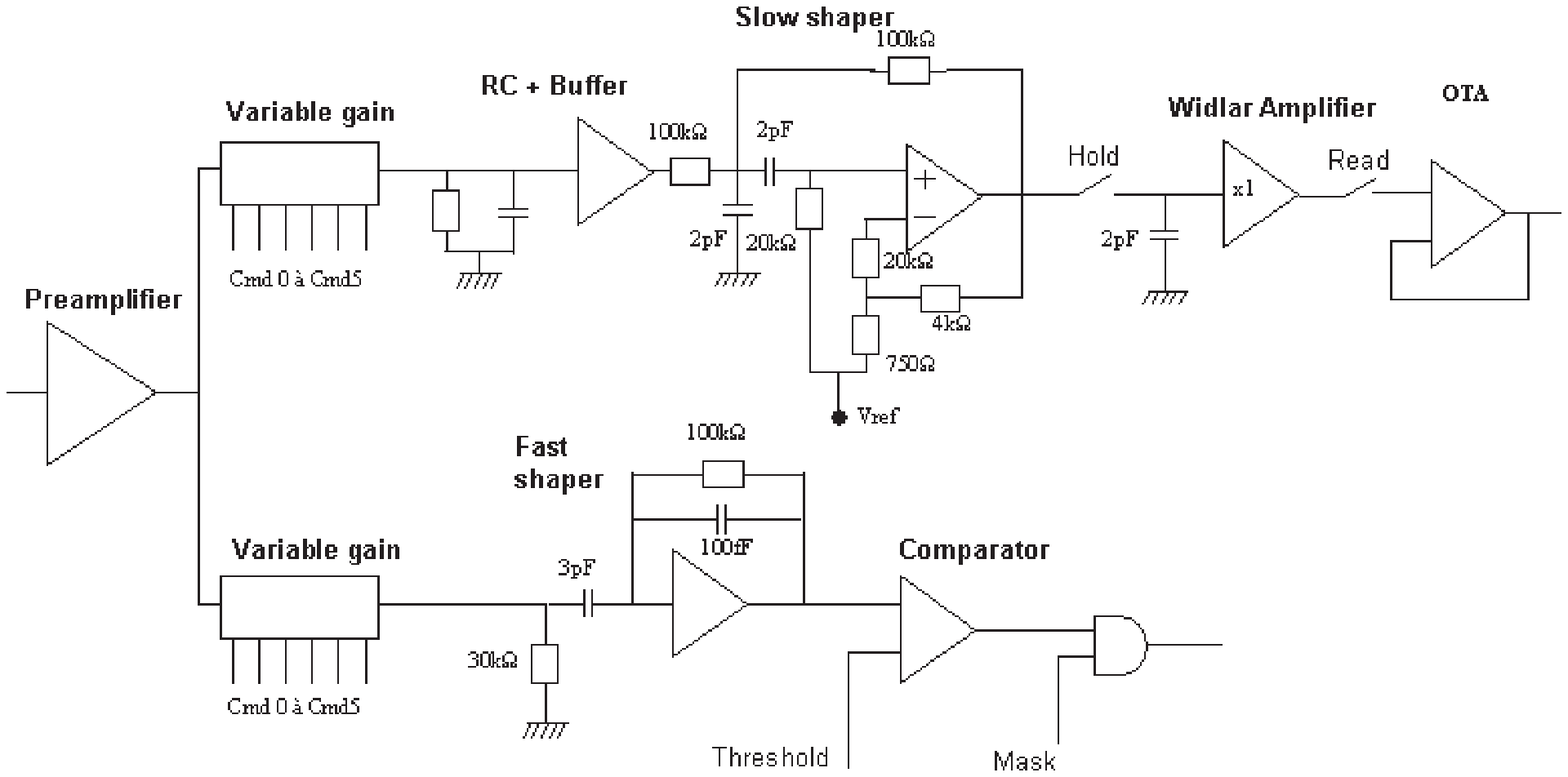,width=10cm}
\caption{\small Architecture of a single channel.}
\label{vers3_global_architecture}
\end{center}
\end{figure}

Each of the 32 channels comprises a low noise variable gain
preamplifier that feeds both a trigger and a charge measurement arms
(Fig.~\ref{vers3_global_architecture}). The auto-trigger includes a
fast shaper followed by a comparator. The trigger decision is
provided by the logical "OR" of all 32 comparator outputs, with a
threshold set externally. A mask register allows disabling
externally any malfunctioning channel. The charge measurement arm
consists of a slow shaper followed by a Track~\&~Hold buffer. Upon a
trigger decision, charges are stored in 2~pF capacitors and the 32
channels outputs are readout sequentially at a 5~MHz frequency, in a
period of 6.4~$\mu$s.

The technology of the chip is AMS BiCMOS
0.8~$\mu$m~\footnote{Technology AutriaMicroSystems (AMS) BiCMOS 0.8~microns,
http://cmp.imag.fr/.}
Its area is about 10~mm$^2$ and it is packaged in a QFP100 case. Its
consumption depends upon the gain correction settings and ranges
between 130 and 160~mW.

The variable gain system is implemented by adding selectable current
mirrors with various areas (2.0, 1.0, 0.5, 0.25, 0.125, and 0.0625).
The activation of the six switches thus allows setting an effective
gain correction ranging from 0 to 3.9. By turning off all current
switches of a channel, the gain is zeroed and the channel is
disabled. For a correction gain set to 1, the preamplifier gain is
found to be 94~mV/pC , or 15~mV/p.e at a PMT gain of $\rm
10^6$, with a rise time of about 30~ns . After amplification, two
copies of the input current are made available to feed both the
trigger and the charge measurement arms. For both arms, the RMS of
the noise corresponds to at most 0.01 p.e.

The fast shaper is directly fed with a mirror output via a 3~pF
capacitance and the signal is integrated in a 0.1~pF charge
amplifier. The integration time constant is 10~ns to produce a fast
signal. A differential input is used to minimize offset dispersion
and to allow a common threshold for the chip with a minimal threshold
spread. The fast shaper is then followed by a comparator, whose
input stage includes a bipolar differential pair in order to
minimize the offset. With a low offset comparator and a high gain in
the shaping just before, a common threshold can be used for all
channels. The trigger decision is defined as the logical ``OR'' of
all comparator's outputs and sets in the charge integration process.

Fast shaper characteristics are a gain of 2.5~V/pC, i.e.,
400~mV/p.e and a peaking time of 10~ns for a preamplifier
gain set to 1. The trigger rise--time only slightly depends on the
input charge magnitude.

The trigger efficiency has been measured as a function of the injected
charge for each individual channel. 100\% trigger efficiency is
obtained for input charge as low as 1/10th of p.e,
independently of the preamplifier correction. The trigger threshold
common to all channels being set externally, the output spread among
the 32 channels has been carefully controlled. It is found to be
around $\rm 0.03$ p.e, an order of magnitude smaller than
the useful threshold level.

The slow shaper has a long peaking time to minimize the sensitivity
to the signal arrival time. The voltage pulse available on the RC
integrator is shaped by a Sallen-Key shaper characterized by a time
constant of $\rm 200~ns$. This corresponds to an average rise--time
of 160~ns with a spread among the 32~channels not exceeding $\rm \pm
4$~ns. In order to minimize pedestal variation from
channel-to-channel slow shaper DC offset dispersion, a differential
input stage has been used.

\begin{figure}[hbt]
\begin{minipage}{.45\linewidth}
\begin{center}
\mbox{\epsfig{file=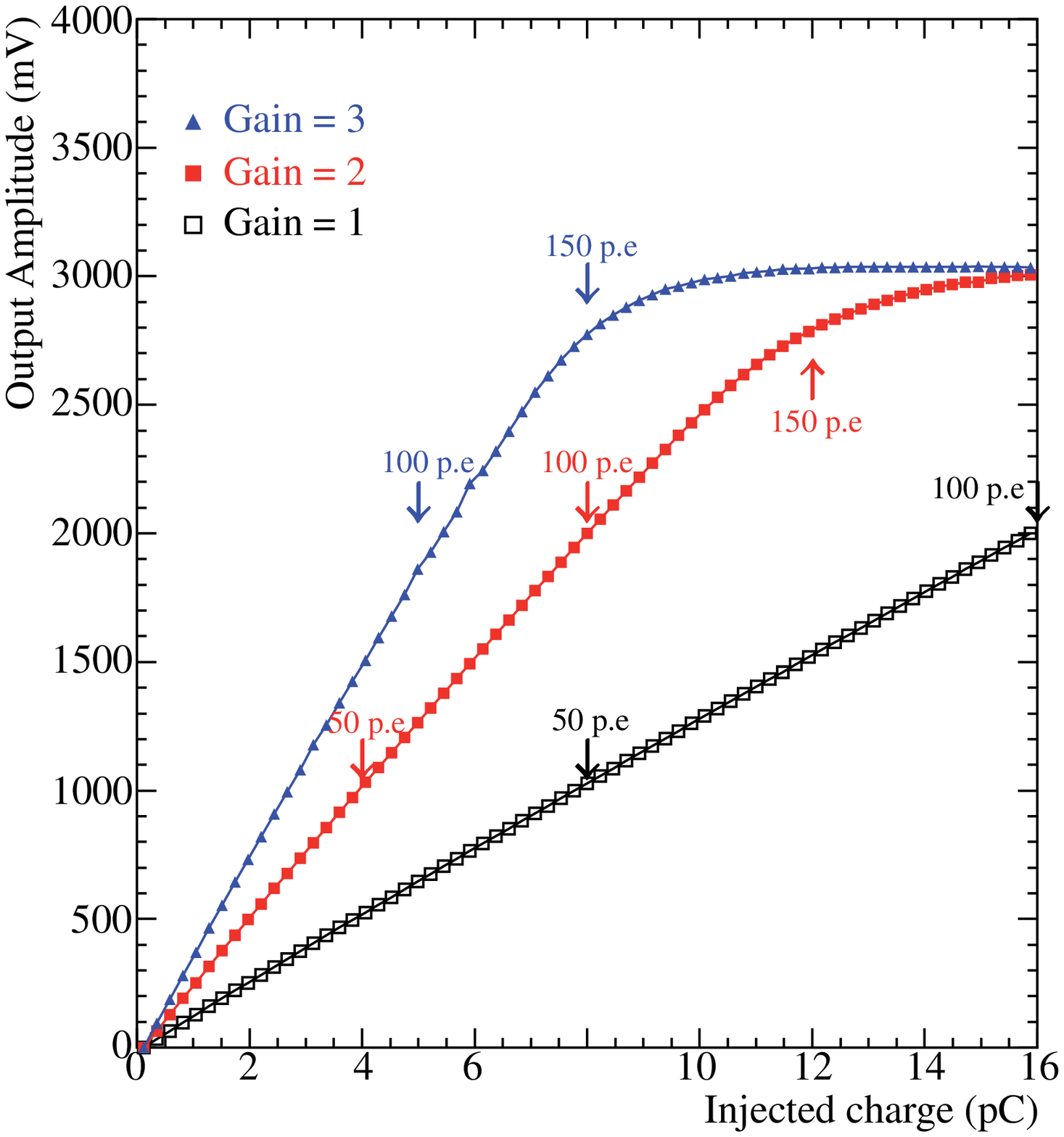,width=7cm}}
\caption{\small Linearity of the charge measurement as function of
the input charge for a gain set at 1, 2 and
3.}\label{vers3_charge_linearity_vs_gain}
\end{center}
\end{minipage} \hspace{1.cm}
\begin{minipage}{.45\linewidth}
\begin{center}
\mbox{\epsfig{file=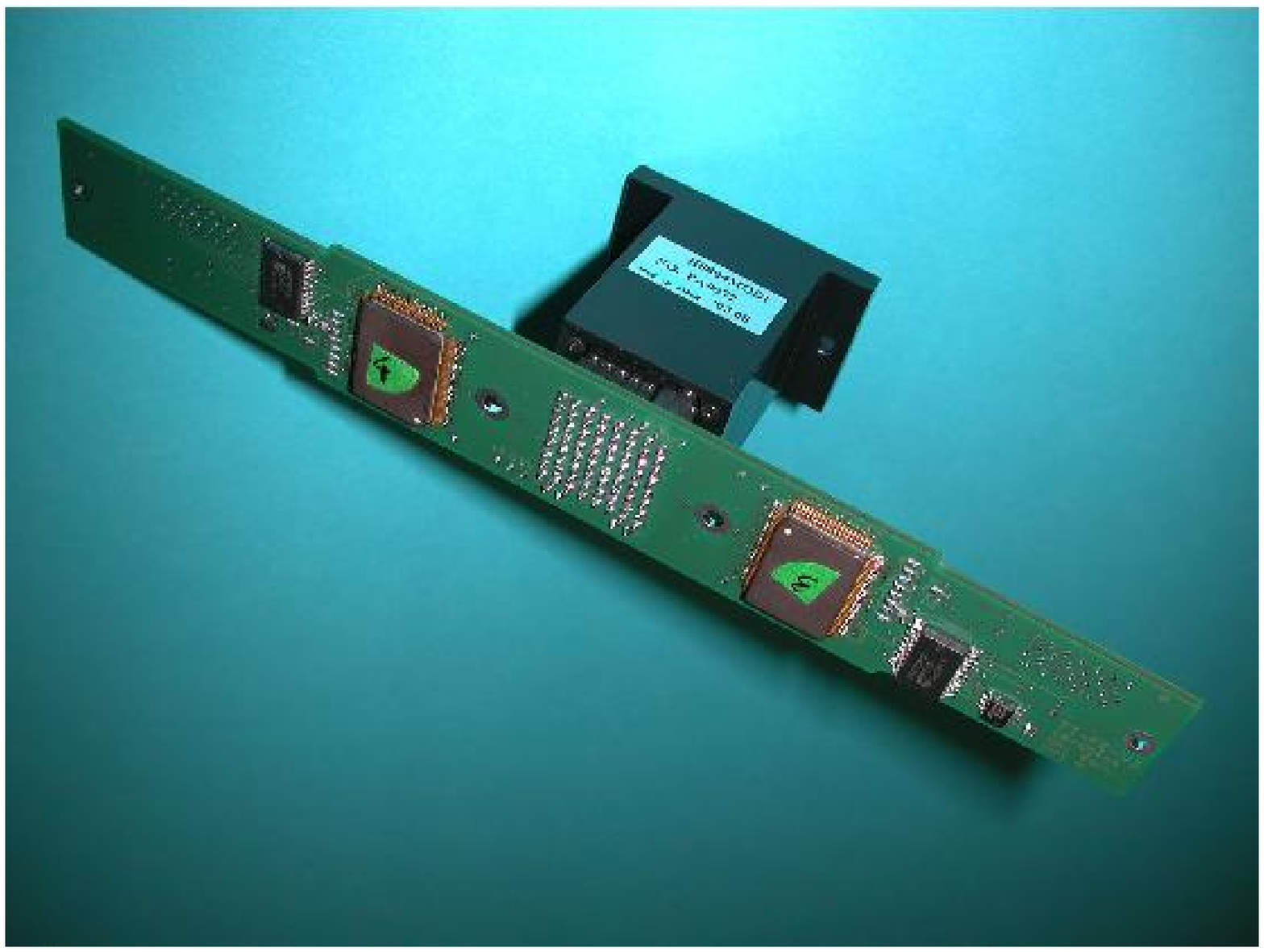,width=7cm}}
\caption{\small The front--end PCB connected to a multianode PMT.}
\label{front_end_PCB}
\end{center}
\end{minipage}
\end{figure}

Upon a trigger decision, all the capacitors are read sequentially
through a shift register made by D-flip-flop. The pedestal level is
$\rm 1.2$~V in average, with a corresponding spread of $\rm\pm
6~mV$. These numbers correspond to less than 1/3rd of
p.e. The signal entering the ADC is in fact the difference
between the multiplexed output and the Channel~0 output. This
channel being disconnected from any input, this technique allows to
make measurements insensitive to pedestal variations common to all
channels caused by, e.g., temperature effects.

The linearity in the charge measurement has been determined for all
channels and found to be better than 2\% over the full range 1-16~pC
for a preamplifier gain of 1, corresponding to 1-100 p.e
(Fig.~\ref{vers3_charge_linearity_vs_gain}).

The noise at the multiplexed output has been measured at 12~fC, i.e.,
0.075 p.e for a preamplifier gain 1, and at 3.7~fC or
0.08 p.e for a maximal gain.

Cross--talk due to the ASIC has been carefully considered. Two main
sources of cross--talk have been identified. A first effect is
interpreted as a coupling between the trigger and the charge
measurement arms and has been determined to be lower than 0.1\%.
The second effect affects the nearest neighbours of a hit channel, where a
cross--talk of the order of 1\% is measured.
This effect is negligible for far channels where a constant
cross--talk component of about 0.2\% is observed.

The front--end board is a 8--layer PCB carrying two ROCs. It is
directly plugged to the PMT, as shown in Fig.~\ref{front_end_PCB}.
The lines from the PMT to the ROC inputs are well separated and
protected from external noise sources by four ground planes in the
PCB. The front--end board contains buffer amplifiers for the
differential charge output signals of the ROC's and logic level
translators for the digital signals. It also carries five
operational amplifiers, four for adapting the output voltage range
of the two DAC's on the DAQ board to the threshold voltage range of
the two ROC's and one for reading the high voltage of the PMT. The
front--end card is connected to the DAQ board by means of two
26--lines miniature flat cables. The ADC's (Analog Devices AD9220,
12 bits) are located on the DAQ board, in order to minimize the
length of the data bus.

%\clearpage
\subsection{Light Injection System} \label{injection}

The light injection system is used to test and monitor all the
electronic channels and the data acquisition system. In each
end--cap, light is injected into the WLS fibres just in front of the
fibre--PMT opto--coupler with the help of LED's
(Kingbright\footnote{Kingbright No. 317-1, Sec. 2 Chung Shan Road
Chung Ho, Taipei Hsien 235 Taiwan.} L-7113PBC), straight PMMA light
guides,  6~mm diameter and 50~mm long, and a white painted diffusive
box (Fig.~\ref{endcap_3D}). The LED's are pulsed from outside the
end--cap by a purposely designed driver activated by an external
trigger.

The gain monitoring of each PMT channel and its associated
electronics can be performed in a very short time in the single
p.e mode and up to about 100 p.e. By pulsing
LED's on one side of the module and reading the signal on the other
side, a possible WLS fibre ageing can be monitored.

The system is designed to provide a rather uniform illumination, within a factor
3, of all the 64 fibres bundled in a $8\times 8$ dense
pattern near the optical window.

The light injection system will regularly be operated during the
whole experiment duration.

\begin{figure}[hbt]
\begin{center}
\epsfig{file=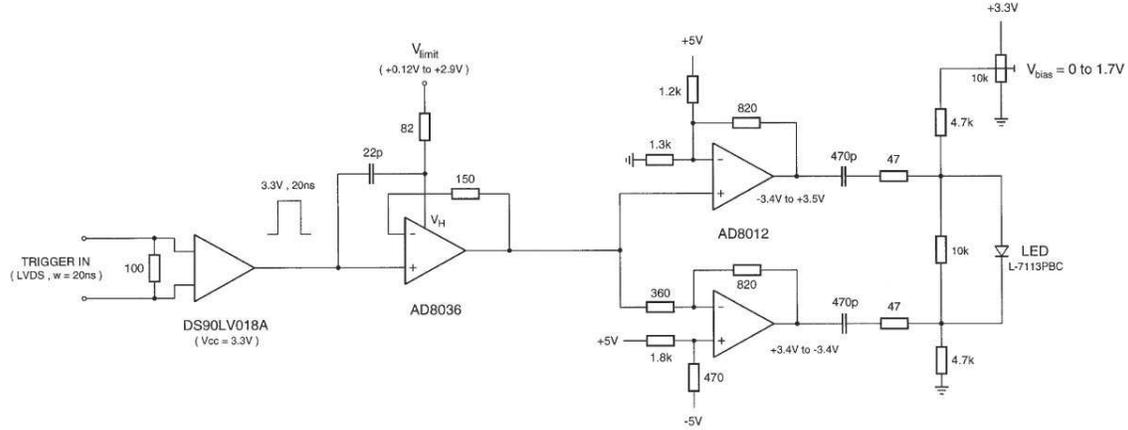,width=15cm}
\caption{\small Simplified circuit diagram of the LED pulser. Only
one of the two output stages for the LED pair of the light
injector is shown.} \label{LED_fig_1}
\end{center}
\end{figure}

The LED pulser (Fig.~\ref{LED_fig_1}) provides fast blue light
pulses with an amplitude range of more than a factor of 200, with a
stability and reproducibility of about one percent at the high and
medium amplitudes and a few percent at the lowest amplitudes. The
circuit, which generates the current pulses for the pair of LEDs of
the light injector, based on standard fast amplifier chips, needs
only $\pm$5~V supplies and fits on a PCB area of $25\times
50$~mm$^2$. The pulser requires a LVDS trigger signal with a width
of 20~ns.

The spread of the absolute light output of 2500 LED's was found to
be within a factor of two. Also the relative light output as a
function of the DAC setting can differ up to a factor of two for two
arbitrary chosen LED's. To improve the uniformity, LED's of similar
performance were chosen to form a pair and adequate filters were
used to equalize the absolute light yield for all pulsing systems.
The spread of the maximum light signal produced by the LED's is then
reduced to about $\pm$15\% for all injectors and the relative light
signals of the LED pairs as a function of the DAC setting will track
within about $\pm$20\%.

\begin{figure}[hbt]
\begin{center}
\epsfig{file=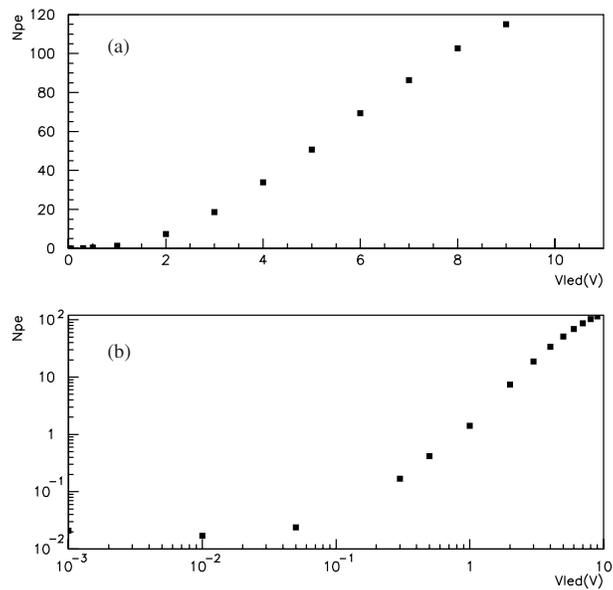,width=8cm}
\caption{\small Response of one channel as a function of the voltage
applied to the LED, a) linear scale showing the linearity at high
voltage, b) log/log scale showing the sensitivity at low voltage.}
\label{injection_lin}
\end{center}
\end{figure}

The variation of the response of one channel as a function of the
control voltage applied to the LED is displayed in Fig.~\ref{injection_lin}.
The non linear response below 20 p.e
makes the amplitude setting at the lower end of the range less
critical. Signals as low as 0.02 p.e can be extracted
from the background.

%\clearpage

%\clearpage
\section{Radioactivity in the Target Tracker} \label{radioactivity}

In this section the radioactivity measurements of the materials used
for the Target Tracker construction are presented. The goal of these
measurements was mainly to ensure that the background caused
by the activity of these materials was low enough not to introduce
significant dead time during the acquisition due to high trigger
rate. Using the results of these measurements, a simulation to
obtain an estimation of  the induced background in the scintillator
strips and measured by the WLS fibres and the PMT's, has been
prepared.

Target Tracker radioactivity has been measured using $\gamma-$ray spectrometry
with an ultra low background germanium detector. The ``$p-type$" 400~cm$^3$ $Ge$
semiconductor is well shielded and located underground at the ``La Vue--des--Alpes"
laboratory in the Swiss Jura Mountains~\cite{vda} under 600~m water--equivalent.
This detector permits to reach a sensitivity of $10^{-10}$~g/g for $U$ and $Th$ and
$10^{-21}$~g/g for $^{60}Co$.

The radioactivity of the materials selected for the construction of the Target Tracker
are summarized in table \ref{resact1}. These materials have
been measured during around one week each.

\begin{table}[hbt]
\begin{center}
\caption{\small Results of the radioactivity measurements.}
\label{resact1}
\begin{scriptsize}
\begin{tabular}{|c||c|c||c||c||c|}
\hline  Materials                                           & \multicolumn{5}{|c|}{Activity (Bq/g)}   \\ \hline
                                                            & \multicolumn{2}{|c||}{$^{238}$U series} & $^{232}$Th series      & \multicolumn{2}{|c|}{} \\ \hline
                                                            & $^{238}$U                               & $^{226}$Ra             & $^{228}$Ra             & $^{137}$Cs or $^{60}$Co & $^{40}$K \\ \hline
Amcrys                                                      &                                         &                        &                        & $^{137}$Cs              &          \\
scintillator                                                & equil.                                  & (1.8$\pm$0.2)10$^{-5}$ &                        &
(3.8$\pm$1.3)10$^{-6}$                                      &                                         \\ \hline
 optical fibre                                              & equil.                                  & (4.0$\pm$0.6)10$^{-4}$ & (4.1$\pm$1.2)10$^{-4}$ &
(6.6$\pm$2.5).10$^{-5}$                                     & (1.2$\pm$0.5)10$^{-3}$                  \\ \hline glue 815 C     & equil.                 & (3.9$\pm$0.8)10$^{-4}$  &          &
                                                            & (3.2$\pm$3.2)10$^{-4}$                  \\ \hline glue hardener  & equil.                 &
 (1.9$\pm$0.3)10$^{-4}$                                     &                                         &                        &                        \\ \hline
   TiO$_2$                                                  &                                         &                        &                        & $^{137}$Cs              &          \\
(R104)                                                      &                                         &                        & (1.7$\pm$0.7)10$^{-5}$ &
(5.9$\pm$2.6)10$^{-6}$                                      &                                         \\ \hline
M 2755                                                      &                                         &                        &                        & $^{60}$Co               &          \\
(adhesive)                                                  & equil.                                  & (1.6$\pm$0.3)10$^{-4}$ &
(3.7$\pm$2.2)10$^{-5}$                                      & (2.3$\pm$2.0)10$^{-6}$
                                                            & (10.0$\pm$9.2)10$^{-5}$                 \\ \hline
sikaflex 221                                                & (5.4$\pm$0.4)10$^{-5}$                  &
(2.4$\pm$0.1)10$^{-3}$                                      & (1.6$\pm$0.3)10$^{-5}$                  &                        &
 (1.2$\pm$0.1)10$^{-4}$                                     \\ \hline
 alum. cover                                                &
 (5.5$\pm$0.7)10$^{-3}$                                     & (1.5$\pm$0.4)10$^{-5}$                  &
 (4.6$\pm$0.2)10$^{-4}$                                     &                                         & (6.1$\pm$3.4)10$^{-5}$ \\
\hline foam (end--caps)                                     & equil.                                  &
 (4.4$\pm$1.5)10$^{-5}$                                     & (1.9$\pm$1.0)10$^{-5}$                  &                        & (8.4$\pm$0.2)10$^{-3}$ \\ \hline
iron                                                        &                                         &                        &                        & $^{137}$Cs              &          \\
end--cap                                                    & equil.                                  & (6.6$\pm$0.7)10$^{-6}$ &
(3.3$\pm$0.9)10$^{-6}$                                      & (1.2$\pm$0.6)10$^{-6}$                  &
(1.9$\pm$0.5)10$^{-5}$                                      \\ \hline
\end{tabular}
\end{scriptsize}
\end{center}
\end{table}

\begin{figure}[hbt]
\begin{center}
\leavevmode \epsfig{figure=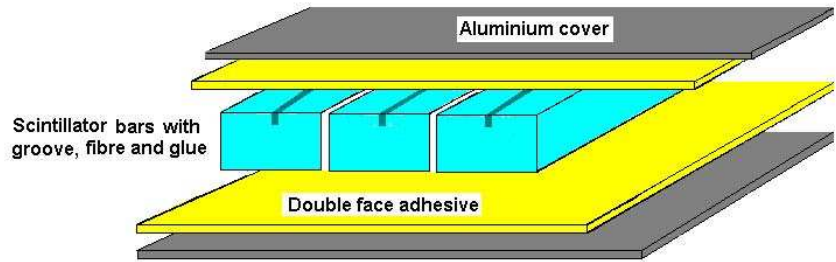,height=4.8cm}
\caption{\small Drawing of the geometry used for the simulation.} \label{geom}
\end{center}
\end{figure}

These measurements have been
used to estimate the signal induced in a scintillator strip. For
this, all the particles ($\alpha$, $\beta$, $\gamma$) coming from
the decay chain of the radioactive elements are simulated for each
component. Then the produced particles are propagated in the Target
Tracker geometry and the energy they deposit  in the scintillator
strip is simulated. Fig.~\ref{geom} shows the geometry used for the
simulation.  3 scintillator strips have been considered in order to
estimate the signal produced in the middle one.

Table \ref{rate} gives an estimation of the rate induced by the
radioactivity of each component and the total rate. The quoted error
only reflects the uncertainty on the radioactivity measurement. The
main contribution is given by the aluminium cover which produces a
total rate of 0.24~Hz. The contributions of the glue and the
fibre are negligible. The total background rate per scintillator
strip induced by the radioactivity of all components is estimated to
0.29~$\pm$0.05~Hz, where the enlarged error includes the
uncertainty on the triggering level which could be different in the
experiment than the one used in the simulation.

\begin{table}[hbt]
\begin{center}
\caption{\small Signal rate estimation induced by each component.} \label{rate}
\begin{scriptsize}
\begin{tabular}{|c||c|c|c|}   \hline
Materials                                                                       & Decay serie & Rate (Hz)    & Total rate (Hz) \\ \hline
Aluminium                                                                       & $^{238}U$   & 0.212$\pm$0.028    &                       \\
                                                                                & $^{232}Th$  & 0.017$\pm$0.001    &                       \\
                                                                                & $^{226}Ra$  & 0.006$\pm$0.0001   &                       \\
                                                                                & $^{40}K$    & 0.005$\pm$0.003    & 0.240 $\pm$0.032      \\ \hline
Central Scintillator strip                                                      & $^{238}U$   & 0.039$\pm$0.005    &                       \\
                                                                                & $^{137}Cs$  & 0.008$\pm$0.003    & 0.047 $\pm$0.008      \\ \hline
Double face                                                                     & $^{238}U$   & 0.002$\pm$0.0003   &                       \\
adhesive                                                                        & $^{232}Th$  & 0.0004$\pm$ 0.0002 &                       \\
                                                                                & $^{60}Co$   & 0.0001$\pm$0.0001  &                       \\
                                                                                & $^{40}K$    & 0.004$\pm$0.004    & 0.0065 $\pm$0.0046    \\ \hline
 Side Scintillator strips                                                       & $^{238}U$   & 0.0006$\pm$0.0001  &                       \\
                                                                                & $^{137}Cs$  & 0.0004$\pm$0.0001  & 0.001 $\pm$0.0002     \\ \hline
Fibre in the central strip                                                      &             & negligible         &                       \\ \hline
Fibres in the side strips                                                       &             & negligible         &                       \\ \hline
Glue in the central strip                                                       &             & negligible         &                       \\ \hline
Glue in the side strips                                                         &             & negligible         &                       \\
\hline \hline  \multicolumn{3}{|c|}{\bf{TOTAL  RATE (s$^{-1}$)}}                &
\bf{0.29 $\pm$0.05}                                                             \\ \hline
\end{tabular}
\end{scriptsize}
\end{center}
\end{table}

%\clearpage
\section{Effect of the magnetic field on PMT's}

The efficiency of a PMT is affected by a strong enough magnetic
field because the Lorentz force modifies the p.e
trajectory. Similarly, the gain is reduced by the effect of the
field on the multiplication process of the secondary electrons.
Studies done by MINOS collaboration and Hamamatsu indicate that the
efficiency of the PMT used in OPERA decreases significantly if the
magnitude of the field perpendicular to the photocathode exceeds 5
Gauss.

\begin{figure}[hbt]
\begin{center}
\epsfig{file=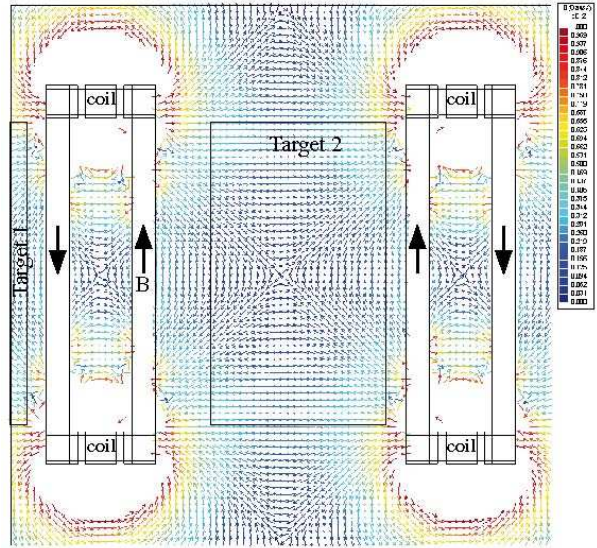,width=8cm}
\end{center}
\caption{\small Geometry of the two spectrometers and magnetic
field map for a range [0,100] Gauss}
\begin{center}
\end{center}
\label{topo_spectro}
\end{figure}

\begin{figure}[hbt]
\begin{center}
\epsfig{file=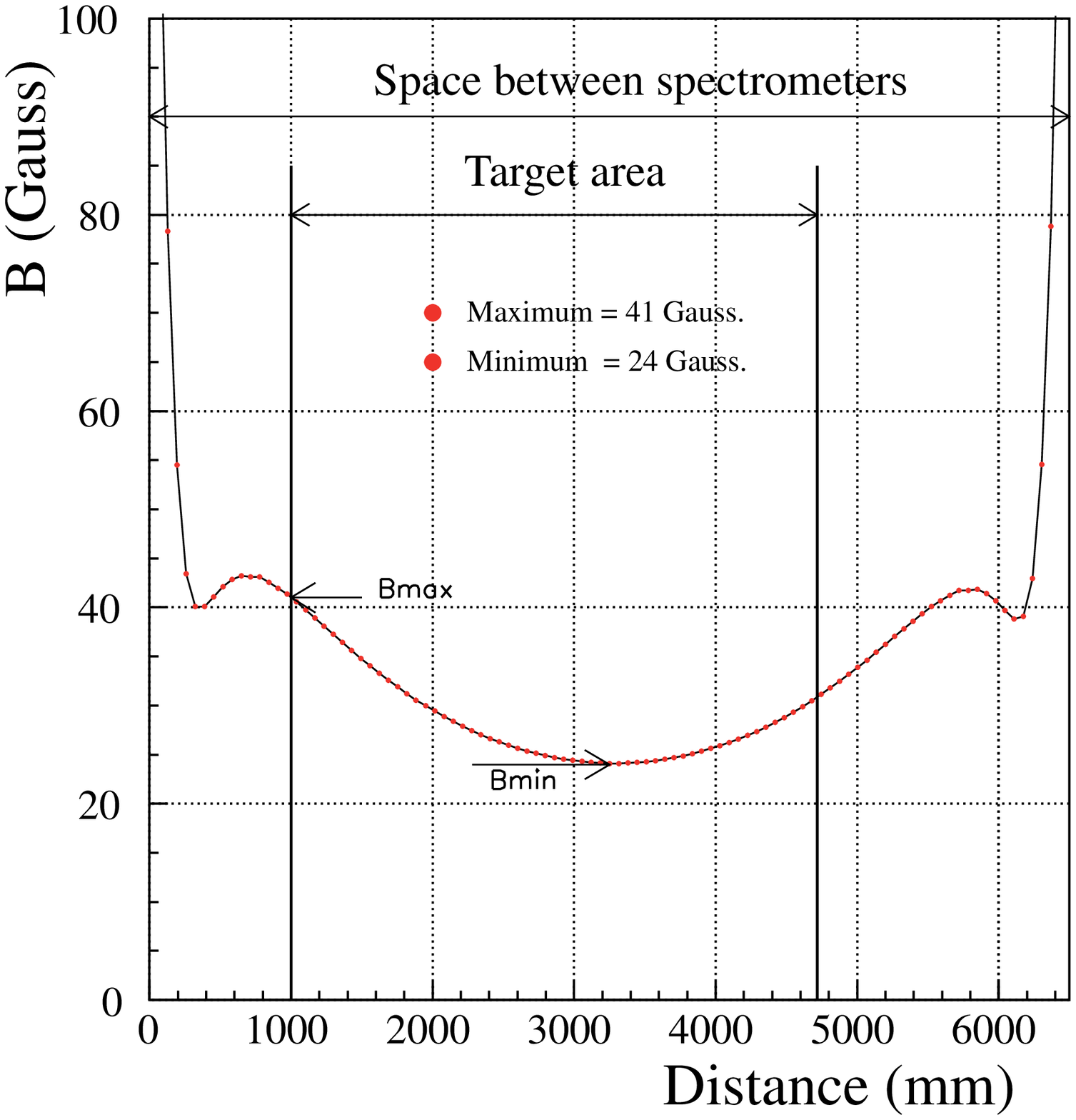,width=7cm}
\epsfig{file=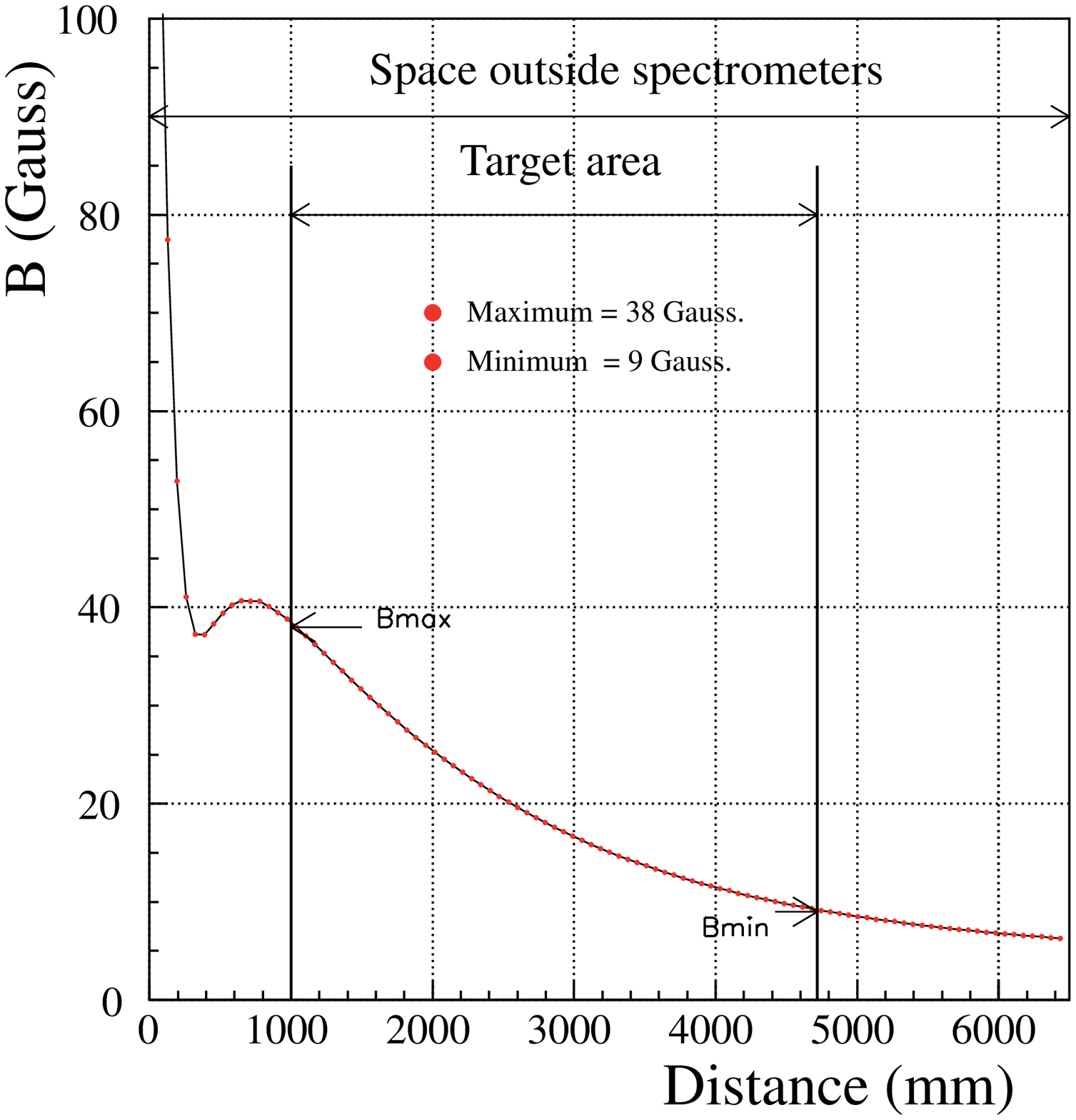,width=7cm}
\caption{\small Simulated magnetic field intensity along the beam axis ($z$)
for PMT row number 3 versus the distance from the two OPERA
magnets.} \label{min_max_definition}
\end{center}
\end{figure}

Simulations done with TOSCA~\cite{tosca}, based on the {\it finite element
method}, and AMPERES~\cite{ampere, baussan}, based on the {\it boundary element
method}, (Fig.~\ref{topo_spectro}), revealed that the magnetic field
at the place where PMT's had to be located was largely exceeding the allowed maximum
value. Fig~\ref{min_max_definition} shows the variation of the
magnetic field along a line parallel to the beam direction and
passing through the series of top PMT's of row 3 (see
Fig.~\ref{wall_schematic} for PMT numbering) where the fringe field
was expected to be maximum. The right--hand plot is for the first
target, upstream of the first spectrometer and the left--hand plot
for the second target, between the two spectrometers. In both cases,
the field varies between 10 and 40 Gauss, showing the necessity to
shield the PMT's.

Evaluations of the effects of a magnetic field on the performance of
the PMT have been performed with a coil surrounding the PMT and its
electronics. The efficiency, the gain and the cross--talk between
channels have been studied by varying the direction and the
magnitude of the field. All 64 photocathode channels were uniformly
illuminated with a blue LED. The ratio
$R_e={N_{pe}(B)}/{N_{pe}(B=0)}$ of the number of p.e
($N_{pe}$) with and without magnetic field is used as a measure of
the efficiency.

\begin{figure}[hbt]
\begin{minipage}{.45\linewidth}
\begin{center}
\mbox{\epsfig{file=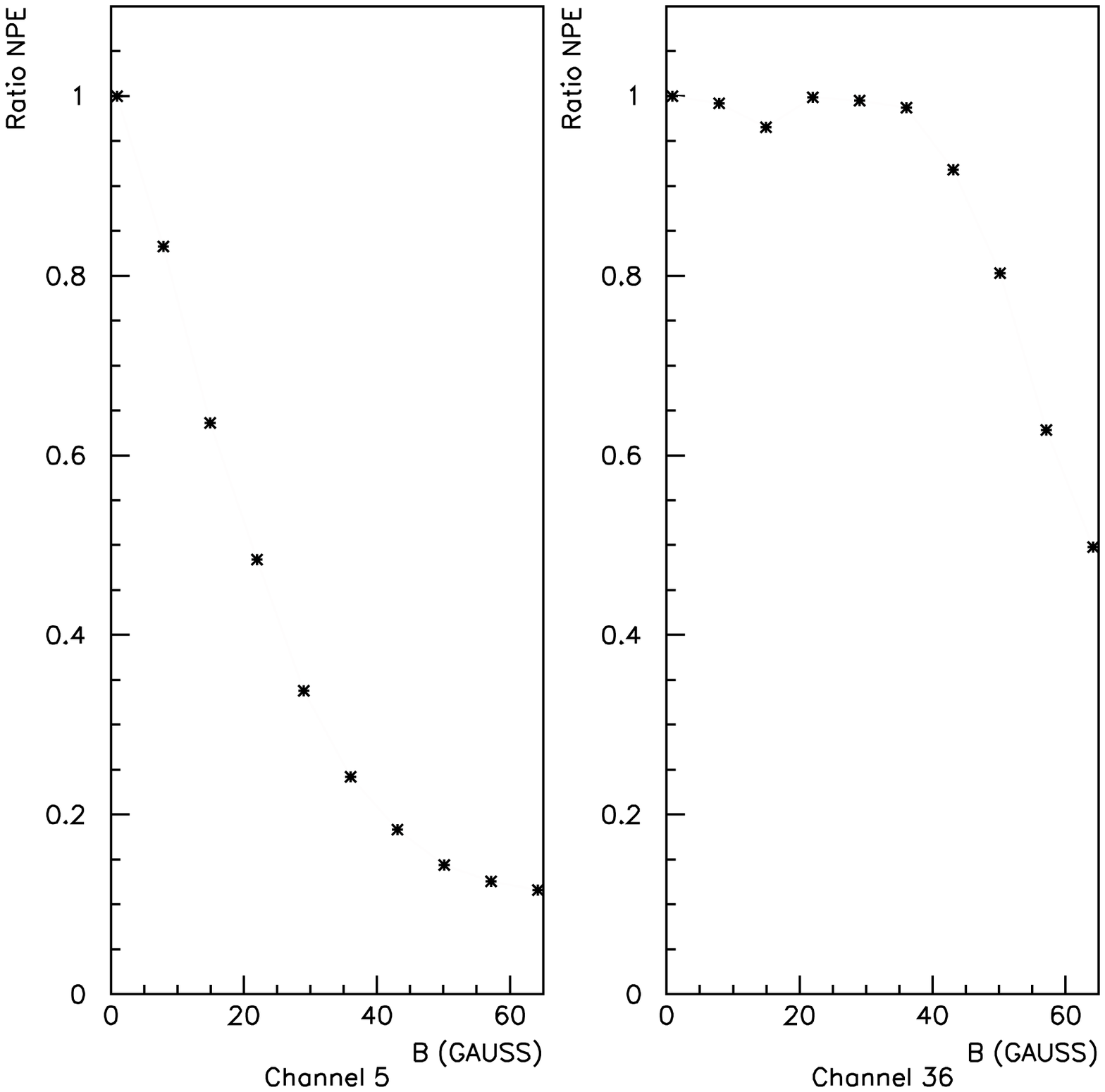,width=7cm}}
\caption{\small Efficiency ratio for channel \#5 (border) and
channel \#36 (central).} \label{efficiency_results}
\end{center}
\end{minipage} \hspace{1.cm}
\begin{minipage}{.45\linewidth}
\begin{center}
\mbox{\epsfig{file=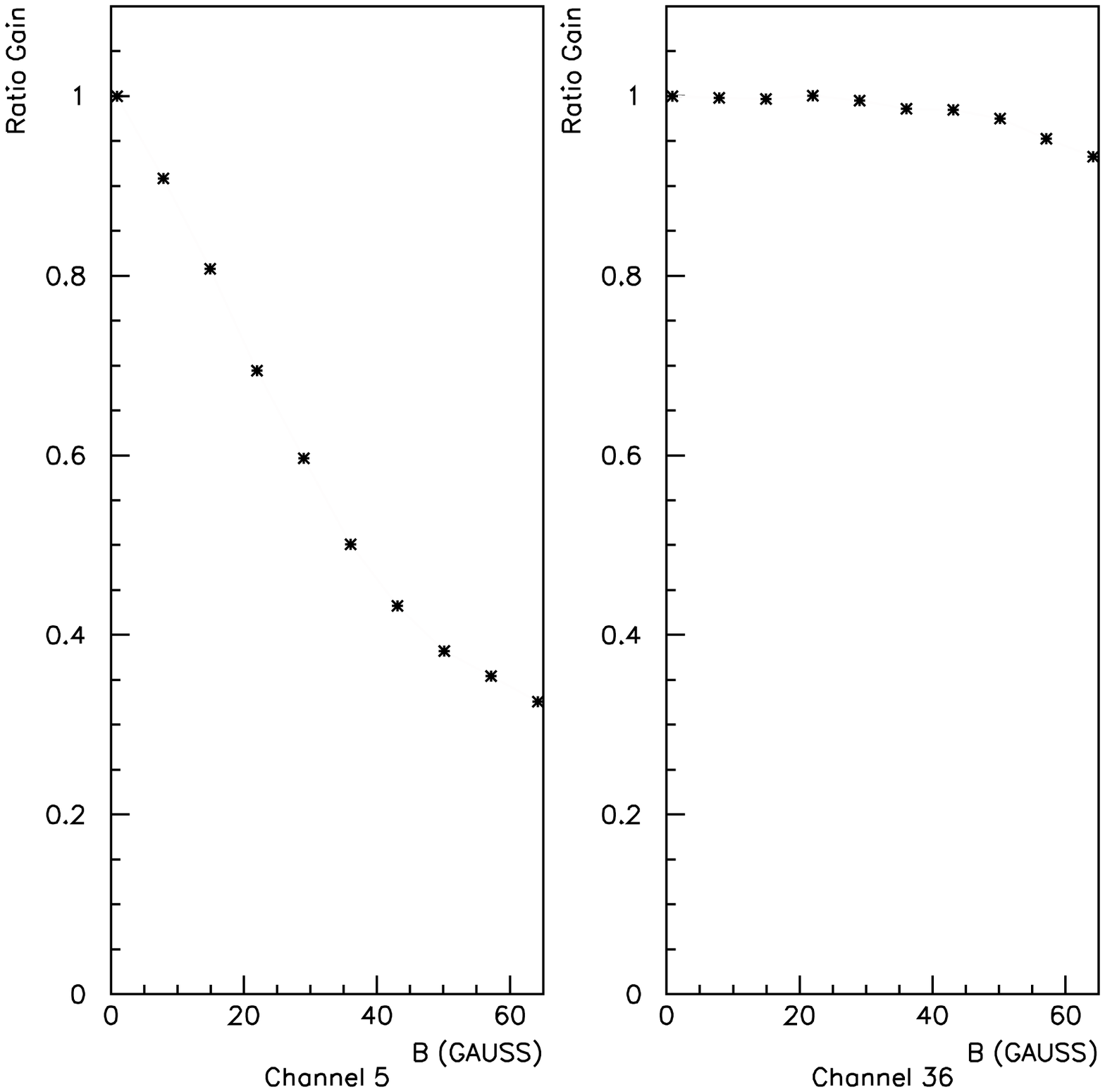,width=7cm}}
\caption{\small Gain ratio for channel \#5 (border) and channel
\#36 (central).} \label{gain_results}
\end{center}
\end{minipage}
\end{figure}

Very different  behaviours are observed  for channels located on the
border and at the centre of the photocathode
(Fig.~\ref{efficiency_results}). The border channels are very
sensitive to the magnetic field while those at the centre show no
significant effect below 30 Gauss. Similar results, though less
dramatic, are obtained for the gain dependence to the field,
evaluated by the ratio $R_G={G(B)}/{G(B=0)}$ of the gains G with and
without magnetic field (Fig.~\ref{gain_results}).

\begin{figure}[hbt]
\begin{center}
\epsfig{file=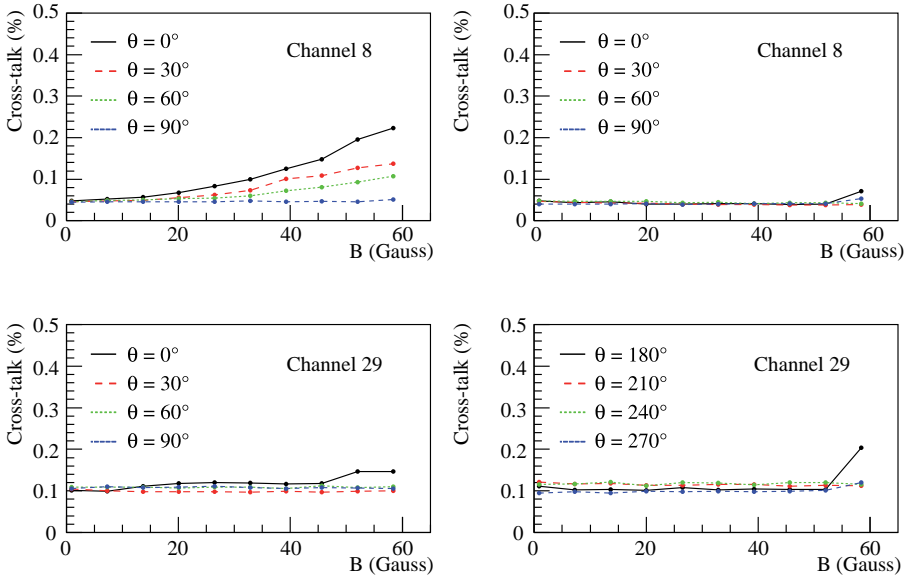,width=12cm}
\caption{\small Cross–-talk level for one channel at the border
(\#8) and one channel at the centre (\#29) of the photocathode for
different magnetic field directions ($\theta=0^\circ$ orthogonal to
the photocathode plane).} \label{crosstalk_results}
\end{center}
\end{figure}

The drop in collection efficiency is expected to be associated with
an increase in the cross--talk between channels, a fraction of the
electrons being deviated towards the neighbouring channels. One
channel in the centre (\#29) and one on the border (\#8) have been
illuminated in turn, varying the direction and the intensity of the
magnetic field. The border channels are significantly affected
whereas the central ones are not and the effect is maximal when
the field is orthogonal to the photocathode
(Fig.~\ref{crosstalk_results}).

\begin{figure}
\begin{center}
{ \epsfig{file=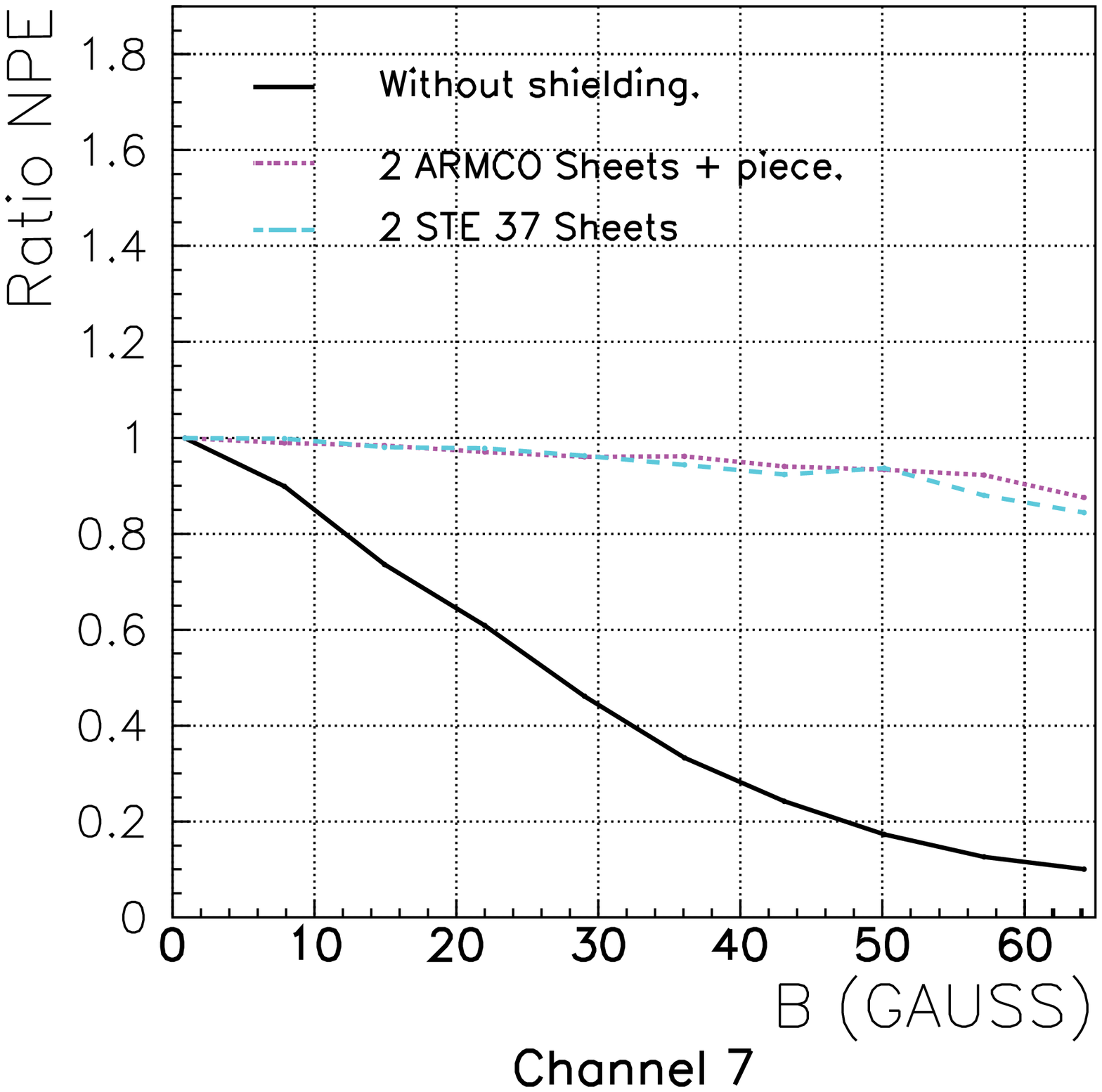,width=7cm}
\hspace{1.cm}
\epsfig{file=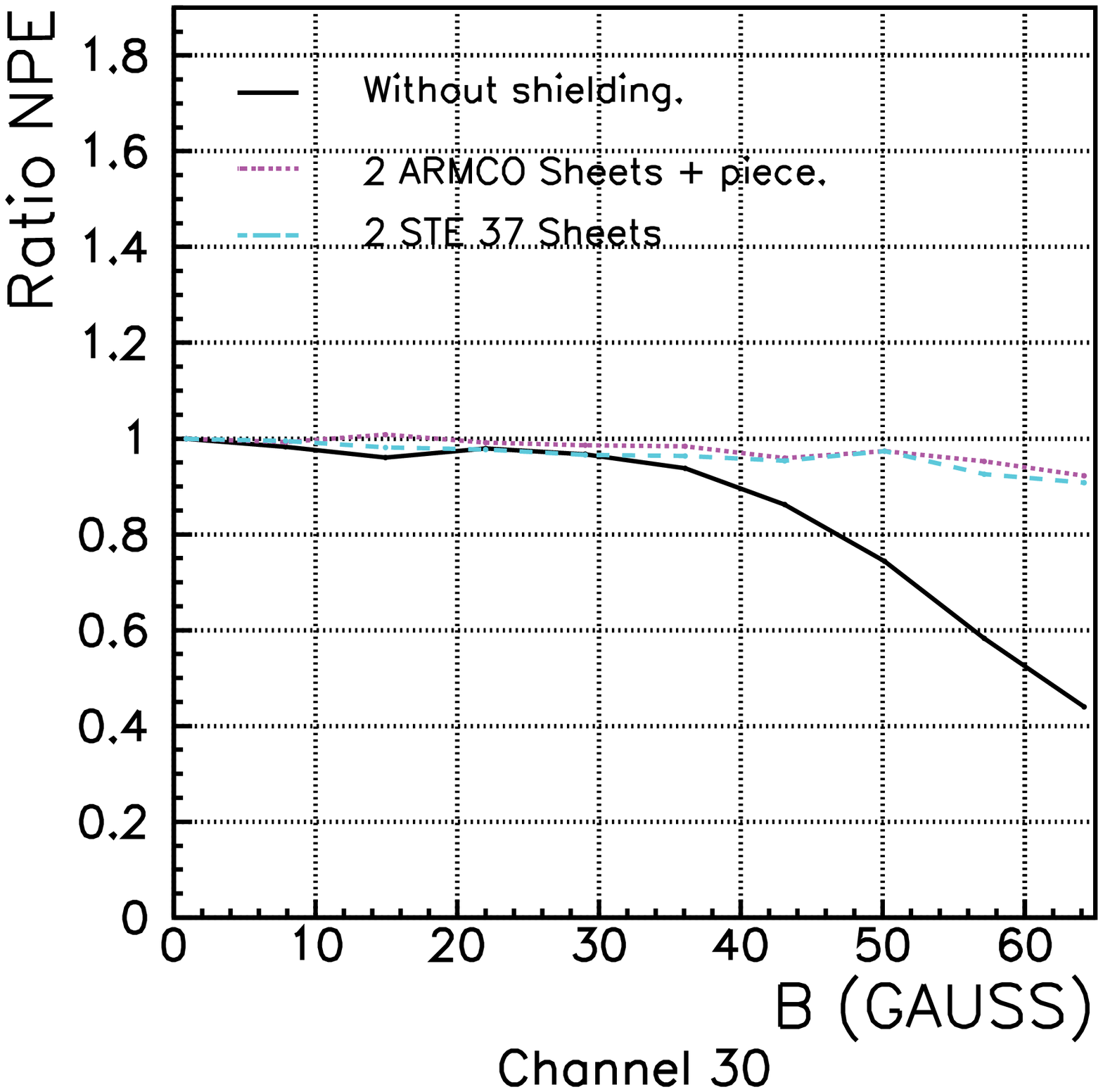,width=7cm} }
\caption{\small PMT efficiency dependence on the magnetic field
orthogonal to the photocathode for no shielding and for the final
shielding using ARMCO and STE 37 steel. Left: channel at the border
(\#7). Right: channel at the centre (\#30).}
\label{armco_ste_shielding}
\end{center}
\end{figure}

Several configurations and materials for magnetic shielding were tested. All the
measurements were done with a field perpendicular to the
photocathode. Mu-metal (main features are high permeability and low
saturation field $B_s=0.8$~T) was tested but discarded because of its
prohibitive price. Adequate shielding was obtained with 0.8~mm STE37
or ARMCO (99.5\% Fe, less expensive than Mu-metal) sheets extending
over the PMT and the fibres optical window and using ARMCO
instead of aluminium for the bar supporting the fibre--PMT
opto--coupler. STE37 was chosen for the covers for reason of cost.
Fig.~\ref{armco_ste_shielding} presents the collection efficiency of
one border and one central PMT channel with and without applying the
chosen shielding. One can see that this shielding is very efficient
for both channels up to a magnetic field of 40~Gauss.

%\clearpage
\section{Construction and Installation} \label{construction}

Two production lines have been set up at IReS--Strasbourg to construct, test and calibrate
a total of 8 modules per week. The construction was done according
to the following steps:

\begin{enumerate}

  \item Groups of 16 scintillator strips were placed on a frame on a
  table equipped with a 16--head glue distribution system.
  The frame was slightly curved along its length, upward at the centre.
  After mixing with the hardener, about 15~g of glue was injected in each groove.
  The fibres, stretched with springs at both ends, were positioned inside the grooves.
  The frame curvature maintained the fibres at the bottom of the grooves by gravity
  during glue polymerization. After polymerization, an aluminized self adhesive
  Mylar band was laid over the grooves to increase the light collection.

  \item The two end--caps were placed on the assembly table on reference
  positions with pins and holes. A double face adhesive was glued on the bottom
  aluminium cover sheet. This sheet was placed on the table, positioning accuracy
  being achieved by inserting at each end the 32 end--caps rivets in the 32 holes of the covers.
  To increase light tightness, black glue was used to assemble the end-caps and the cover together.

  \item The top double face adhesive protection was removed.
  Thick strings were placed transversally on the adhesive every 40~cm to prevent the contact
  of the strips with the adhesive before final positioning. The 64 strips were
  positioned in the module. Each strip was accurately positioned at one end, alternatively
  left and right, by inserting the corresponding end--cap rivet in its hole.
  Insuring the proper alignment of the two external strips with guides provided the overall geometrical accuracy.

  \item All fibres ends were inserted in their nominal holes in the fibre/PMT opto--couplers.
  This very delicate manipulation was executed prior to gluing the strips to allow their
  replacement in case of accidental damaging of the fibres.

  \item The strings were removed and the strips were glued on the double face adhesive.
  The top aluminium cover with its double face adhesive was put in place and pressed against the strips.
  Again, its holes and the end-caps rivets insured its proper
  positioning.

  \item The two light injection systems were mounted in the end--caps.

  \item Black glue was injected in the small tank of the opto--couplers.
  The holes of the opto--couplers being  0.08~mm larger than the fibre diameter,
  the glue filled the holes and slightly expended to the outside.

  \item The end--caps covers were sealed with black glue.
  For mechanical protection and to insure light tightness, an aluminium
  ribbon reinforced with carbon fibres was glued on the two long edges of the module.
  The ribbon was itself further protected by a 0.1~mm U--shape stainless steel foil.

  \item After polymerization of the glue in the opto--coupler, the module was placed on
  its side in a vacuum box. Two elastic membranes applied pressure on the two aluminium
  cover sheets to complete the assembly.

  \item The opto--coupler faces were polished with a Diamond Head countersink.

  \item The two PMT's, their front--end electronic boards, DAQ boards and pairs of LED's
  for the light injection system were mounted in the end-caps and connected.

\end{enumerate}

Using this procedure, only a negligible number of fibres have been broken at one end ($<0.2$\%) during the module construction and after
the strip gluing.

The modules were transported to Gran Sasso in special boxes of 8 modules, 4 boxes per truck, equipped with shock absorbers.
The 62 Target Tracker walls were mounted in situ in groups of 8, in parallel with the insertion inside the detector of previously mounted walls.
The alignment of the inserted walls was performed using the Leica TDA5005 theodolite. Marks on both end--caps of the modules were used as
reference points.

%\clearpage
\section{Module tests and calibration} \label{calibration}

The light tightness of the modules was tested using PMT's dynode 12
which is the OR of all the PMT channels. The same measurements
repeated in the Gran Sasso underground laboratory under much reduced
cosmic ray and ambient radioactivity background give a counting
rate of the order of 50~Hz/channel in the absence of light leaks.

For energy calibration, each module was placed on a vertical scanning table
equipped with two electron spectrometers (see \ref{scintillators}) that may irradiate
simultaneously any two points of its surface. The nominal high
voltage values determined during the PMT acceptance tests previously described,
were applied to the tested module PMT's.
The correction factors equalizing all the PMT's channels were set in
the front--end chips.

\begin{figure}[hbt]
\begin{center}
\epsfig{file=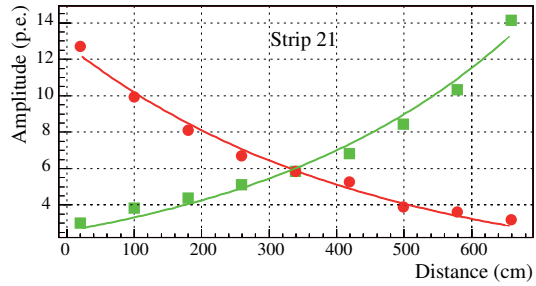,width=7cm}
\caption{\small Number of p.e observed at each end of a
strip during the calibration of a Target Tracker module versus the distance from the photodetectors.}
\label{calib_curve}
\end{center}
\end{figure}

\begin{figure}[hbt]
\begin{minipage}{.45\linewidth}
\begin{center}
\mbox{\epsfig{file=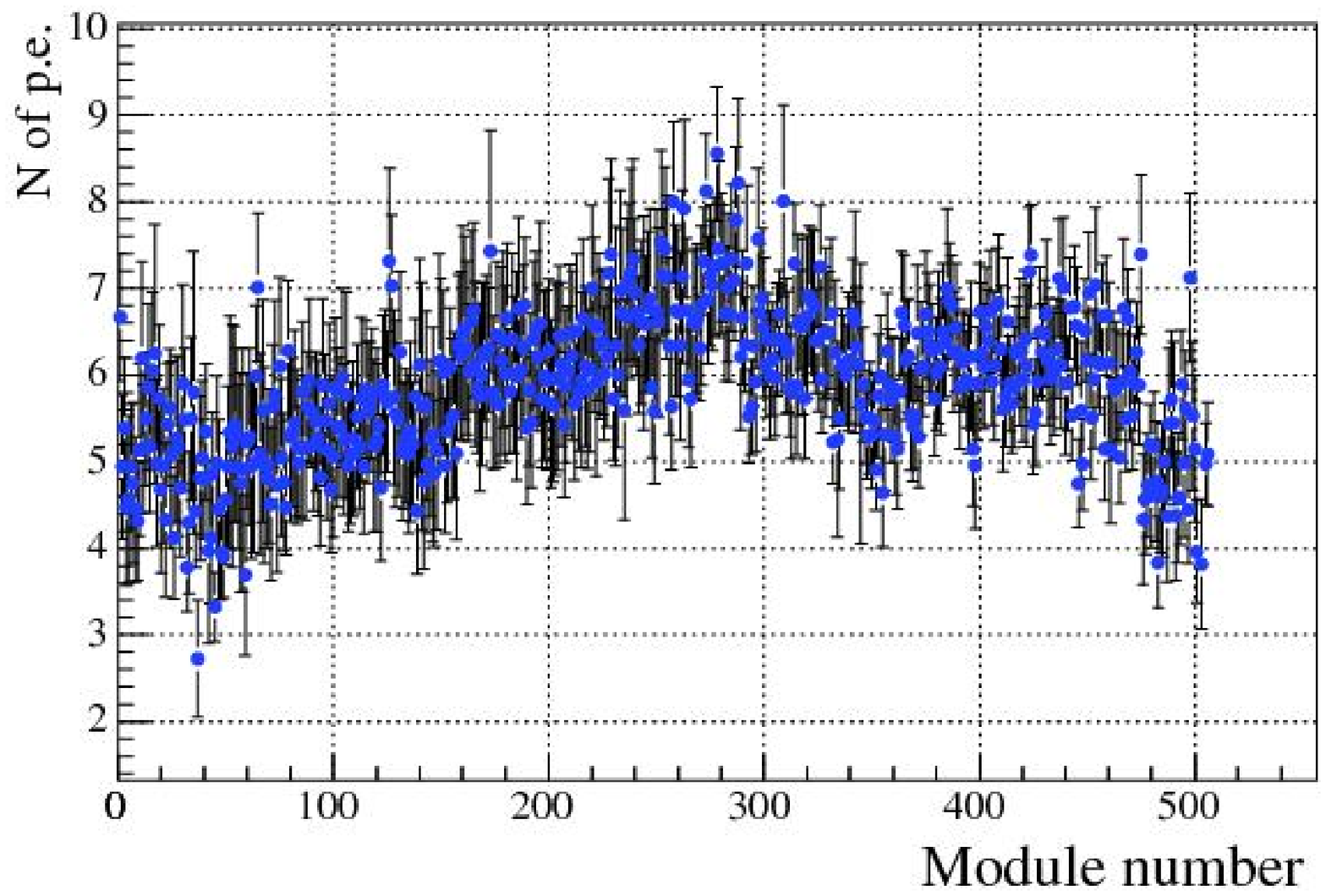,width=7cm}}
\caption{\small Mean value of the number of p.e observed
at the middle of the strips versus the module number (or versus the
scintillating strip production time, the production lasted 2
years).} \label{all_calibration}
\end{center}
\end{minipage} \hspace{1.cm}
\begin{minipage}{.45\linewidth}
\begin{center}
\mbox{\epsfig{file=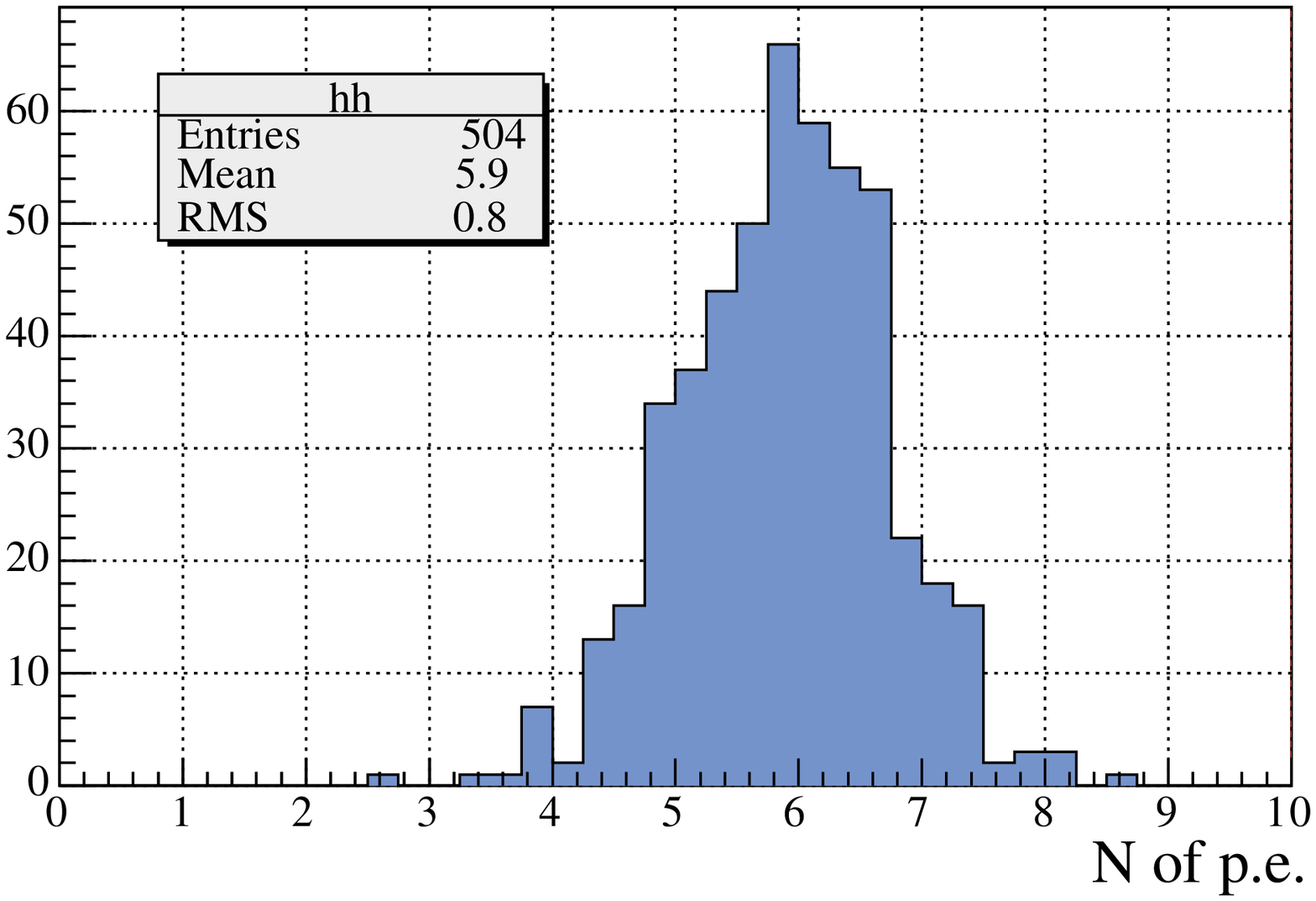,width=7cm}}
\caption{\small Distribution of the mean value of the number of
p.e observed at the middle of the scintillating strips
averaged over the Target Tracker module ($y$--projection of
Fig.~\ref{all_calibration}).} \label{mean_calibration}\end{center}
\end{minipage}
\end{figure}

Two scanning tables driven by the acquisition program have been used
during the production period. Measurements were taken at 9 points
uniformly distributed along the length of each strip (see example
Fig.~\ref{all_calibration}). All distributions of the number of p.e versus the distance have
been fitted by an exponential distribution. The parameters are
stored in a data base for further use during the analysis
to reconstruct by calorimetry the energy of OPERA events.
Fig.~\ref{all_calibration} shows the mean number of p.e
observed at the middle of the scintillator strips versus the strips
production time. Fig.~\ref{mean_calibration} is the marginal
distribution of the former. The mean number of p.e is 5.9.
Essentially, all values are well above the lower limit of 4 required in the specifications.

%\clearpage
\section{Aging}

The properties of the glues are well known as they have been used by other neutrino
experiments (NEMO3~\cite{nemo}, MINOS). Several glue samples
have been followed for more than 7 years. The strength of the double
face adhesive is guaranteed by the production and selling companies
(MACTAC\footnote{MACtac Europe, Boulevard Kennedy, B-7060 Soignies,
Belgium} and VARITAPE\footnote{Varitape, NV Frankrijkstraat, 5 9140
Temse, Belgium}) not to vary for at least 10 years. According to
tests done by the AMCRYS--H company, an acceptable 10\% decrease in
the light output of the strips is expected over 10 years. Kuraray
fibres are also widely used and tested by high energy experiments in
much more severe conditions than in OPERA, as for example by the LHC ATLAS~\cite{atlas}
experiment where no ageing effects have been reported. No ageing
is expected for the multi-channel PMT's.

During construction, a piece of 100~mm was cut from one every 8th
scintillator strip. 63 optically isolated pieces glued together
form a strip of the same length as a regular strip. 64 such strips
made of 7936 pieces were assembled into a module. The module has
been placed on a scanning table and each piece is irradiated in turn
by an electron spectrometer following the procedure developed for the
calibration of all Target Tracker modules. Any of the 7936 pieces
has its light output measured every 25 days by continuously running
the system, thus allowing a follow--up of the quality of
the scintillator strips and of the glue used for fibres. Natural
radioactivity in the underground laboratory is also being considered
to measure the time dependence of the light output by dedicated
measurements considering the shape of the corresponding energy distribution
measured by the Target Tracker.

%\clearpage
\section{Acknowledgments}

We acknowledge the support of the funding agencies IN2P3--France,
FNRS/IISN--Belgium, SNF--Suiss and JINR-Russia. We would like to thank all
private companies which have collaborated with our institutes in
developing and providing materials for the OPERA Target Tracker
construction. We are grateful to the OPERA collaboration for the support provided.
Finally, we would like to thank all technicians non--authors of this paper who have
helped us all these last years.
%\clearpage

\end{document}